\documentstyle[prb,aps,eqsecnum,multicol,psfig,array]{revtex}
\def\wide#1#2{
\end{multicols}
\widetext
\noindent
\if#1t
\else
    \raisebox{9pt}[0in][0.0in]
    {$\rule{3.4in}{0.4pt}\rule{0.4pt}{6pt}$\hspace{3.6in}}       
\fi
#2
\if#1b
\else
    \raisebox{-9pt}[0in][0.0in]
    {\hspace{3.55in}$\rule{0.4pt}{6pt}\rule[6pt]{3.5in}{0.4pt}$}
\fi
\begin{multicols}{2}
\noindent
}

\begin{document}
\title{Localization length at the resistivity minima of the
quantum Hall effect}
\author{M. M. Fogler, A. Yu. Dobin, 
        and B. I. Shklovskii}

\address{Theoretical Physics Institute, University of Minnesota,
116 Church St. Southeast, Minneapolis, Minnesota 55455}

\maketitle

\begin{abstract}

The resistivity minima of the quantum Hall effect arise due to the
localization of the electron states at the Fermi energy, when it is
positioned between adjacent Landau levels. In this paper we calculate
the localization length $\xi$ of such states at even filling factors
$\nu = 2 N$. The calculation is done for several models of disorder
(``white-noise,'' short-range, and long-range random potentials). We find
that the localization length has a power-law dependence on the Landau level
index, $\xi \propto N^\alpha$ with the exponent $\alpha$ between one
and $\frac{10}{3}$, depending on the model. In particular,
for a ``white-noise'' random potential $\xi$ roughly coincides
with the {\em classical\/} cyclotron radius. Our results are in
reasonable agreement with experimental data on low and moderate mobility
samples.

\end{abstract}

\begin{multicols}{2}

\section{Background and results}
\label{intro}

The appearance of narrow resistivity peaks separated by deep minima is a
defining feature of the quantum Hall effect (QHE).~\cite{Prange} The
explanation of such a dependence of $\rho_{xx}$ on the magnetic field
$B$ is based on the idea of localization. The states at the Fermi energy
are localized at almost all $B$ except for a few discrete values $B_N$
where the Fermi energy is at the center of $N$th Landau level (LL). Near
such special values the localization length $\xi$ is believed to diverge,
\begin{equation}
      \xi \propto |{B - B_N}|^{-\gamma},
\label{divergent xi}
\end{equation}
where $\gamma$ is a critical exponent.  The analytical calculation of
$\gamma$ is a notoriously difficult problem.  (Numerical methods give
$\gamma = 2.35 \pm 0.03$, see Ref.~\onlinecite{Huckestein}).  At the same
time, the calculation of $\xi$ away from the critical region turns out to be
much simpler.  Such a calculation is the subject of the present paper.  As a
demonstration of the method, we calculate $\xi$ at discrete values of $B$,
$B \simeq (B_N + B_{N + 1}) / 2$.  They correspond to the minima of
$\rho_{xx}$ in the transport measurements.

Generally speaking, the definition of $\xi$ is not unique. In this paper
we will adopt the following one,
\begin{equation}
\frac{1}{\xi} =  -\lim\limits_{r\to\infty}\frac{1}{2 r}
\left\langle \ln|\psi(\bbox{r})|^2 \right\rangle,
\label{decay}
\end{equation}
where $\psi(\bbox{r})$ is the wave function of the state at the Fermi
level. The averaging is assumed to be done over the disorder realizations.

Our definition of the localization length is chosen to represent an
experimentally measurable quantity. Indeed, it is well known that
transport at sufficiently low temperatures proceeds via the
variable-range hopping. In its turn, the hopping conduction is
determined by the {\em typical\/} decay rate of the tails of the wave
functions. Definition~(\ref{decay}) relates $\xi$ precisely to this
typical rate.

Let us further elaborate on this point. The bulk of low-temperature
experimental data on the quantum Hall
devices~\cite{Ebert,Briggs,Goldman} can be successfully fit to the
following dependence of $\rho_{xx}$ on the temperature $T$,
\begin{equation}
                \rho_{xx} \propto e^{-\sqrt{T_0 / T}},
\label{T_one_half}
\end{equation}
which can be interpreted~\cite{Polyakov} in terms of the variable-range
hopping in the presence of the Coulomb gap.~\cite{ES} In this theory
$T_0$ is directly related to $\xi$ defined by Eq.~(\ref{decay}),
\begin{equation}
          k_B T_0 = {\rm const}\, \frac{e^2}{\kappa \xi},
\label{T_0}
\end{equation}
where $e$ is the electron charge and $\kappa$ is the dielectric constant
of the medium. Using Eq.~(\ref{T_0}), one can extract the dependence of
$\xi$ on $B$ from the low-$T$ transport measurements in the
straightforward way. (We will discuss experimentally
relevant issues in more detail in Sec.~\ref{conclusions}.)

In this paper we calculate $\xi$ using a model where the
disorder is described by a Gaussian random potential $U(\bbox{r})$ with
the two-point correlator
\begin{equation}
\langle U(\bbox{r}_1) U(\bbox{r}_2)\rangle = C(|\bbox{r}_1 - \bbox{r}_2|).
\label{C_r}
\end{equation}
We will assume that function $C(r)$ becomes small at $r$ larger than
some distance $d$ and that $C(r)$ does not have any other characteristic
lengths. The rms amplitude of the potential, $\sqrt{C(0)}$, will be
denoted by $W$. We will assume that $W$ is much smaller than $E_F$, the
Fermi energy. The electron-electron interaction is ignored at this
stage.

\begin{figure}
\centerline{
\psfig{file=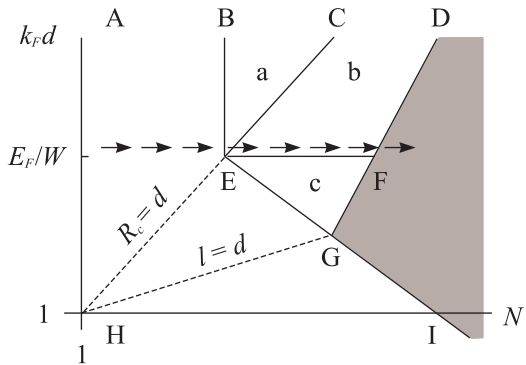,width=2.7in,bbllx=180pt,bblly=396pt,bburx=560pt,bbury=660pt}
}
\vspace{0.1in}
\setlength{\columnwidth}{3.2in}
\centerline{
\caption{The parameter space of the problem with a {\em monoscale\/}
random potential [Eqs.\ (\protect\ref{C_r}) and (\protect\ref{C_q})].
The entire phase space is divided into regions with different dependence
of $\xi$ on $N$ and $k_F d$. The boot-shaped region AHIGEB is described
by Eq.~(\protect\ref{xi_short}), the shaded region to the right of the
line DFGI by Eqs.~(\protect\ref{ansatz}) and~(\protect\ref{Drude}), the
region BEGFD by Eq.~(\protect\ref{xi_milnikov}), and finally, the region
below the line HI by Eq.~(\protect\ref{xi_white}). The arrows show the
``trajectory,'' traced by a ``standard'' sample (see definition in
Sec.~\protect\ref{conclusions}) as the magnetic field decreases.
\label{diagram}
}}
\end{figure}

To facilitate the presentation of our results we would like to introduce
the phase diagram shown in Fig.~\ref{diagram}. The vertical axis stands
for the dimensionless parameter $k_F d$, where $k_F$ is the Fermi
wave-vector of the two-dimensional electron gas, $k_F = \sqrt{2 \pi n}$,
$n$ being the electron gas density. The horizontal axis is the LL index
$N \simeq (k_F l)^2 / 2$, where $l = \sqrt{\hbar c / e B}$ is the
magnetic length. The Fermi level is assumed to be at the midpoint
between the centers of $N$th and $N + 1$st LLs. The axes are
in the logarithmic scale. The ratio $W / E_F$ is assumed to be fixed.

%
%

Several lines drawn in Fig.~\ref{diagram} divide the phase plane into
the regions with different dependence of $\xi$ on $B$. Let us explain
the physical meaning of these lines. The line BEGI is the line where the
densities of states of neighboring LLs start to overlap as $N$ increases
($B$ decreases). Thus, to the right of this line the density of states
at the Fermi level is practically equal to its zero field value. We will
refer to this region of the parameter space as the region of overlapping
LLs. To the left of the line BEGI only the tails of the neighboring LLs
reach the Fermi level and the density of states is much smaller than at
zero field. This region will be called the region of discrete LLs. The
equation, of the line EGI is (cf. Refs.~\onlinecite{Ando,Raikh,Laikhtman})
\begin{equation}
{\rm EGI}:\:  k_F d \sim \frac{1}{N} \left(\frac{E_F}{W}\right)^2.
\label{EGI}
\end{equation}
Another line in Fig.~\ref{diagram}, GFD, separates the regions of
different dynamic properties. To the left of this line the guiding
centers of the cyclotron orbits would perform the regular drift along
certain closed contours. This phenomenon has been dubbed ``classical
localization'' in Ref.~\onlinecite{Fogler_chaos}. To the right of the line
GFD (shaded sector in Fig.~\ref{diagram}) the motion of the guiding center
is diffusive (on not too large length scales). The equation
for the line GFD has been derived in Ref.~\onlinecite{Fogler_chaos},
\begin{equation}
{\rm GFD}:\: k_F d \sim N \left(\frac{W}{E_F}\right)^{2 / 3}.
\label{GFD}
\end{equation}

In the diffusive region the calculation of $\xi$ reduces to the
calculation of the ``classical''
conductivity $\sigma_{xx}$ by means of the ansatz
(for discussion and bibliograthy see Refs.\onlinecite{Huckestein}
and~\onlinecite{Fogler_chaos})
\begin{equation}
 \xi \propto \exp
\left[\pi^2 \left(\frac{h}{e^2} \sigma_{xx}\right)^2\right],\quad
 \sigma_{xx} \gg \frac{e^2}{h}.
\label{ansatz}
\end{equation}
The ``classical'' $\sigma_{xx}$ is to be calculated by virtue of the
Einstein relation, i.e., as a product of quantum density of states and
the {\em classical\/} diffusion coefficient. The physical mechanism of
the localization in this region is the destructive interference of the
classical diffusion paths. The calculation of the ``classical''
$\sigma_{xx}$ to the right of the line GFD and in the logarithmically narrow
sector to its left (where $\sigma_{xx}$ is still larger than
$\frac{e^2}{h}$) has been done in Ref.~\onlinecite{Fogler_chaos} in some
detail. As one can see from Eqs.~(\ref{EGI}) and~(\ref{GFD}), the studied region
corresponds to rather long range of the random potential, $k_F d > (E_F
/ W)^{2 / 3}$. However, Eq.~(\ref{ansatz}) applies for smaller values of
parameter $k_F d$ as well. That is as long as we stay to the right of
the line GI. In the entire shaded sector to the right of DFGI
the motion is diffusive. The corresponding ``classical''
$\sigma_{xx}$ is given by the usual Drude-Lorentz formula
\begin{equation}
          \sigma_{xx} = \frac{\sigma_0}{1 + (\omega_c \tau)^2}.
\label{Drude}
\end{equation}
Equations (\ref{ansatz}) and (\ref{Drude}) enable one to calculate $\xi$
up to a pre-exponential factor. In this sector $\xi$ is exponentially
large.

As $N$ decreases and the boundary DFGI of the diffusive region is
crossed, the ``classical'' $\sigma_{xx}$ rapidly falls off. Above the point
G this is brought about by the aforementioned ``classical
localization''; below the point G it is caused by the rapid decrease of the
density of states at the Fermi level. Already slightly to the left of
the line DFGI the ``classical'' $\sigma_{xx}$ becomes much less than
$\frac{e^2}{h}$ and the ansatz~(\ref{ansatz}) loses its domain of
applicability. On the physical level, the nature of the particle motion
changes: the diffusion is replaced by quantum tunneling. This paper is
devoted to the calculation of $\xi$ in the tunneling regime
(unshaded area in Fig.~\ref{diagram}).

One has to discriminate between the tunneling in the case of overlapping
LLs and in the case of discrete LLs. The former is realized in the
region above the line BEGFD. The idea of the derivation of $\xi$ in this
regime belongs to Mil'nikov and Sokolov~\cite{Milnikov} who applied it
to the lowest Landau level.~\cite{Shklovskii_84} The argument goes as
follows. In the described regime the density of states near the Fermi
level is high. On the quasiclassical level such states can be thought of
as a collection of close yet disconnected equipotential contours, along
which the particle can drift according to the classical equations of
motion. Nonzero $\xi$ appears as a result of the quantum tunneling
through the classically forbidden areas between adjacent contours. The
localization length is determined by the spatial extent of relevant
equipotential contours and by the characteristic tunneling amplitude.

Reference~\onlinecite{Milnikov} has been criticized in literature for,
e.g., neglecting the interference effects. In our opinion this criticism
is unjustified. The authors of Ref.~\onlinecite{Milnikov} have
clearly indicated the domain of applicability of their theory. It can be
verified that within this domain the amplitude of tunneling between the
pairs of contours is small; hence, the probability of return to the
initial point after at least one tunneling event is also small. In this
case the interference phenomena can be safely ignored (cf.
Ref.~\onlinecite{Huckestein}).

The case of high Landau levels requires some modifications to the
original method of Mil'nikov and Sokolov. The details are given in
Sec.~\ref{percolation}. We have found that the region of the phase space
bounded by the line BEGFD consists, in fact, of three smaller regions
(see Fig.~\ref{diagram}) with different dependence of $\xi$ on $N$:
\begin{mathletters}
\label{xi_milnikov}
\begin{eqnarray}
\displaystyle k_F \xi &\sim& \frac{N^{10 / 3}}{k_F d}
\left(\frac{W}{E_F}\right)^{7 / 3}
\label{xi_I}\\
\displaystyle \mbox{} &\sim& \frac{N^{5 / 2}}{(k_F d)^{1 / 6}}
\left(\frac{W}{E_F}\right)^{7 / 3}
\label{xi_II}\\
\displaystyle \mbox{} &\sim& N^{5 / 2} (k_F d)^{1 / 2}
\left(\frac{W}{E_F}\right)^3
\label{xi_III}
\end{eqnarray}
\end{mathletters}
(the equation labels match the region labels in Fig.~\ref{diagram}). 
The boundary EC, which separates the regions I and II, is given by
\begin{equation}
                 {\rm EC}:\: k_F d \sim N.
\label{EC}
\end{equation}
The variety of different subregimes in the region BEGFD appears
because of an interplay among three important length scales of the
problem: the correlation length $d$ of the random potential, the cyclotron
radius $R_c = (2 N + 1) / k_F$, and the percolation length $\xi_{\rm
perc}$ (the typical diameter of the relevant equipotential contours).
In this connection note that Eq.~(\ref{EC}) is simply $R_c \sim d$.

Compared to such a variety, the situation in the region of discrete LLs
(AHIGEB) is very simple: the dependence of $\xi$ on $N$ is given by a
single formula. Suppose that the Fourier transform $\tilde{C}(q)$ of the
correlator $C(r)$ [see Eq.(\ref{C_r})] has the form
\begin{equation}
\tilde{C}(q) = \tilde{C}(0)\,\exp\!\left[-\frac{1}{\beta}(q d)^\beta\right],
\quad \beta > 1,
\label{C_q}
\end{equation}
then $\xi$ is given by
\begin{equation}
\xi \simeq \left(\frac{2\beta}{\beta - 1} \ln \frac{\hbar\omega_c}{W}
           \right)^{-\frac{\beta - 1}{\beta}}
           \frac{2 l^2}{d}.
\label{xi_short}
\end{equation}
The logarithmic factor neglected, this can be written as
\begin{equation}
                  \xi \sim \frac{R_c}{k_F d}.
\label{xi_short II}
\end{equation}
Formula~(\ref{xi_short}) has previously appeared (for $\beta = 2$) in
the work of Raikh and Shahbazyan.~\cite{Raikh_xi} These authors
considered the case of the lowest LL ($N = 0$) but suggested that it is
also valid for $R_c \ll d$, i.e., within the knife-shaped region AHEB.
We demonstrate that Eq.~(\ref{xi_short}) is in fact valid in a much
larger domain. The differences between this work and
Ref.~\onlinecite{Raikh_xi} are outlined in the end of Sec.~\ref{long}.

The only part of the phase diagram we have not discussed yet is the area
of ``white-noise'' potential. It is located below the line HI, i.e.,
$k_F d = 1$. The corresponding formula for $\xi$ is
\begin{mathletters}
\label{xi_white}
\begin{eqnarray}
\displaystyle \xi &=& \frac{2 R_c}{\cal L},\quad {\cal L} \ll 2 N + 1,
\label{xi_white_N}\\
\displaystyle &=& \frac{l}{\sqrt{\cal L}},\quad {\cal L} \gg 2 N + 1,
\label{xi_white_0}\\
\displaystyle {\cal L} &\simeq& \ln
\left(\frac{\hbar\omega_c l}{W d}\right),
\end{eqnarray}
\end{mathletters}
which matches Eq.~(\ref{xi_short}) at $k_F d \sim 1$. Previously,
formula~(\ref{xi_white_0}) has been obtained by Shklovskii and
Efros~\cite{Shklovskii_83} and also by Li and Thouless~\cite{Li} for the
lowest LL, $N = 0$.

Neglecting the logarithmic factor, we can write Eq.~(\ref{xi_white_N})
in a remarkably simple form,
\begin{equation}
                           \xi \sim R_c.
\end{equation}
Remarkably, the quantum localization length $\xi$ is
determined by a purely classical quantity, the cyclotron radius!

The basic idea used in the derivation of Eqs.~(\ref{xi_short})
and~(\ref{xi_white}) is to study not the tunneling of the particle
itself but the tunneling of the guiding center $\bbox{\rho}$ of its
cyclotron orbit. For definiteness, consider the tunneling in the
$y$-direction. The effect of the magnetic field can be modelled by means
of the effective ``magnetic'' potential,
\begin{equation}
      U_m(y) = \frac{m \omega_c^2 (y - \rho_y)^2}{2},
\label{magnetic barrier}
\end{equation}
acting on the particle. The classical turning points for this type of
potential are at the distance $R_c$ from $\rho_y$. Therefore, if
$\rho_y$ does not change its position, the longest distance that the
particle can travel without getting under the magnetic barrier is $2
R_c$. And since the barrier increases with $y$, the suppression of the
wave function, which starts beyond this distance, is faster than a
simple exponential.

In the absence of the random potential, $\rho_y$ is a good quantum
number; however, if the external potential is present, it can
scatter the particle, which would cause a change in the guiding center
position. Such a ``scattering-assisted'' tunneling modifies the overall
decay of the wave function~\cite{Shklovskii_82,Shklovskii_83,Li} 
from the super-exponential to the plain exponential one,
\[
            \psi(0, y) \sim e^{-y / \xi}.
\]

Denote a typical displacement of the guiding center after one scattering
act by $\Delta\rho_y$. The physical picture of tunneling depends on the
relation between $\Delta\rho_y$ and $2 R_c$.

The case $\Delta\rho_y > 2 R_c$, which is typically realized at the
lowest LL, has been studied previously in
Refs.~\onlinecite{Shklovskii_83,Li,Shklovskii_82}. In this case the
tunneling involves the propagation under the magnetic barrier. Note that
the barrier itself no longer increases as $y$ squared, which would be with
$\rho_y = {\rm const}$ [see Eq.~(\ref{magnetic barrier})]. After a
series of displacements of the guiding center, the barrier acquires a
saw-tooth shape. In this regime the under-barrier suppression is an
important factor in the overall decay of the wave function.

In contrast to the lowest LL, at high LLs ($N \gg 1$) where $R_c$ is
large, the inequality of the opposite sense, i.e., $\Delta\rho_y \leq 2
R_c$, is typically realized. In this case the particle does not
propagate under the magnetic barrier at all! However, the tunneling a
distance $y \gg \Delta\rho_y$ requires a large number $M \sim y /
\Delta\rho_y$ of the scattering acts. The amplitude of each act is
proportional to $W$ and is inversely proportional to a large energy
denominator $E = E_F - \hbar\omega_c(N + \frac12)$. At the resistivity
minima of the QHE, which we are mainly interested in, $E = \hbar\omega_c
/ 2 \gg W$ (discrete LLs), which implies that the typical scattering
amplitude is small and that the wave function decays exponentially with
$y$ even though the electron never propagates under the magnetic barrier
(this argument is simply a verbal representation of the locator expansion). 
  
\begin{figure}
\vspace{0.1in}
\centerline{
\psfig{file=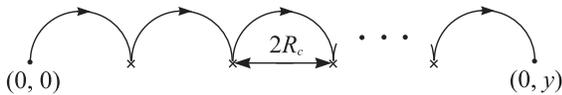,width=2.9in,bbllx=56pt,bblly=410pt,bburx=452pt,bbury=474pt}
}
\vspace{0.1in}
\setlength{\columnwidth}{3.2in}
\centerline{\caption{
The optimal tunneling trajectory of an electron in the random potential of
the ``white-noise'' type. Crosses symbolize the scattering acts. The 
direction of the $y$-axis is from the left to the right.
\label{Olympics}
}}
\vspace{0.1in}
\end{figure}

The case of the ``white noise'' potential is quite illuminating in this
respect. The optimal tunneling path is sketched in Fig.~\ref{Olympics}.
The optimization is based roughly on the requirement that each
scattering event should displace the guiding center by the largest
possible distance without placing the particle under the magnetic
barrier. Clearly, this distance is equal to $2 R_c$, which makes the
tunneling trajectory look like a classical skipping orbit near a hard
wall, see Fig.~\ref{Olympics}.

%
%

Let us estimate the localization length corresponding to this optimal
path. Since $\Delta\rho_y \simeq 2 R_c$, the number of the scattering
events needed to travel the distance $y$ is $M \simeq y / (2 R_c)$. As
discussed above, after each event the wave function decreases by a
factor of the order of $W / (\hbar\omega_c)$. Hence, the overall
suppression factor is $(\hbar\omega_c / W)^{-M}$, which means that $\xi
\sim 2 R_c / \ln(\hbar\omega_c / W)$ in agreement with
Eq.~(\ref{xi_white_N}). This derivation will be done more
carefully in Sec.~\ref{short}.

Formula~(\ref{xi_short}) can be derived in a similar way. After each
scattering event the wave function decreases by a factor of the order of
$f = (W / E) \exp[-(1 / 2 \beta) (q d)^\beta]$ where $q = \Delta\rho_y /
l^2$ is the typical wave vector absorbed in the scattering act. The
total suppression factor after propagating the distance $y$ is of the
order of $f$ to the power $y / \Delta\rho_y$. Optimizing this
suppression factor with respect to $\Delta\rho_y$, one arrives at
Eq.~(\ref{xi_short}). A detailed derivation will be done in
Sec.~\ref{short}.

The paper is organized as follows. Section~\ref{single} is devoted to
general considerations and qualitative derivation of
Eq.~(\ref{xi_short})~and~(\ref{xi_white}). In Sec.~\ref{short} this
derivation is made more rigorous assuming that the random potential is
short-range (or ``white-noise''). In Sec.~\ref{long} we consider the
long-range potential in the regime of discrete LLs. The approach is
different from that of Sec.~\ref{short} but the final result,
Eq.~(\ref{xi_short}), is the same as for the short-range case. In
Sec.~\ref{percolation} we consider the case of overlapping LLs. The
variety of regimes in Eq.~(\ref{xi_milnikov}) is explained with the help
of results developed in the field of statistical
topography.~\cite{Isichenko} Finally, in Sec.~\ref{conclusions} we
compare our results with available experimental data for moderate
mobility samples and propose a method to perform the measurements
with modern high-mobility devices.

\section{General considerations}
\label{single}

Following the overwhelming majority of papers in the field, we will take
advantage of the single Landau level approximation. In this
approximation the original Hilbert space is truncated to the functions,
which belong to the particular ($N$th) Landau level. It is conventional
to choose the orthonormal set of functions
\begin{eqnarray}
\displaystyle \psi_n(x, y) &=& \frac{e^{\frac{i}{\hbar}\rho_n x}}{\sqrt{L_x}}
\Phi_N (y + \rho_n),\label{basis}\\
\displaystyle \Phi_N(y) &=& \frac{1}{\sqrt{2^N n!\,l\pi^{1/2}}} e^{-y^2 / 2 l^2}
H_N\left(\frac{y}{l}\right),\label{Phi_N}\\
\rho_n &=& \frac{2\pi l^2}{L_x} n, \quad n = 0, 1, \ldots
\label{rho_n}  
\end{eqnarray}
to be our basis states. Here $L_x$ is the $x$-dimension of the system and
$H_N(z)$ is the Hermite polynomial. Such functions are the
eigenfunctions of the Hamiltonian
\begin{equation}
H = \frac{(\bbox{p} + \frac{|e|}{c} \bbox{A})^2}{2 m} + U(\bbox{r})
\label{Hamiltonian}
\end{equation}
in the Landau gauge, $\bbox{A} = (By, 0, 0)$, provided
there is no external potential ($U \equiv 0$).

The single Landau level approximation works well if the cyclotron
frequency $\omega_c$ is the fastest frequency in the problem. This is
obviously the case for the discrete LLs, i.e., in the region AHIGEB in
Fig.~\ref{diagram}. It is less trivial and it was demonstrated in
Ref.~\onlinecite{Fogler_chaos} that the inter-LL transitions are
suppressed in the region above the line BEGFD as well. When such
transitions are neglected, the guiding center coordinates, $\rho_y = y -
(v_x / \omega_c)$ and $\rho_x = x + (v_y /\omega_c)$ become the
only dynamical variables in the problem.

Since the random potential is assumed to be isotropic, so is the
ensemble averaged decay of the wave functions. With the above choice of
the basis, however, it is convenient to study such a decay in the
$y$-direction: from the point $(0, 0)$ to the point $(0, y)$.

As the next step we notice that the guiding center coordinates satisfy
the commutation relation
\[
                    \left[\rho_y, \rho_x\right] = i l^2.
\]
Thus, $\rho_y$ plays the role of the canonical coordinate while the
quantity $(\hbar / l^2) \rho_x$ is the canonical momentum. It is
therefore natural to use the $\rho_y$-representation for the wave
functions. For example, in this representation wave
functions~(\ref{basis}) become delta-functions. In general, the
transformation rule between the two representations, $\psi(x, y)$ and
$\phi(\rho_y)$, is given by the formula,
\begin{equation}
\psi(x, y) = \int \frac{d \rho_y}{\sqrt{2 \pi l^2}}\, \Phi_N(y - \rho_y)\,
\phi(\rho_y)\, e^{-\frac{i}{l^2} x \rho_y}.
\label{transform}
\end{equation}

Definition~(\ref{decay}) of the localization length can also be written
in terms of the electron's Green's function $G$,
\begin{equation}
\frac{1}{\xi} =  -\lim\limits_{y\to\infty}\frac{1}{y}
          \left\langle \ln|G(0, y; E)| \right\rangle.
\label{decay_Green}
\end{equation}
On the other hand, Eq.~(\ref{transform}) leads to the following Green's
function transformation rule for $G(0, y; E)$,
\begin{equation}
\displaystyle G = \int\!
\frac{d \rho_1 d \rho_2}{2 \pi l^2}\,
\Phi_N(\rho_1) \, \Phi_N(\rho_2)\,
G_\rho(\rho_1, \rho_2 + y; E),
\label{transform_Green}
\end{equation}
where $G_\rho(\rho_1, \rho_2; E)$ is the Green's function in the guiding
center representation (as in Sec.~\ref{intro} the energy $E$ is
referenced to the Landau level center, and so we are interested mostly
in the case $|E| = \hbar\omega_c / 2$).

Green's function $G_\rho(\rho_1, \rho_2; E)$ satisfies the Schr\"oedinger
equation (with the delta-function as a source)
\wide{m}{
\begin{equation}
G_\rho(\rho_1, \rho_2; E) = \frac{\delta(\rho_1 - \rho_2)}{E}
+ \frac{1}{L_x} \sum_{q_x} U_0(\tilde{q}_x, \rho_1 + \frac12 q_x l^2 )\,
G_\rho(\rho_1 + q_x l^2, \rho_2; E),
\label{Dyson}
\end{equation}
}
where the tilde indicates the Fourier transform over the corresponding
argument. The quantity $U_0$ has the meaning of the random potential
averaged over the cyclotron orbit (cf. Refs.~\onlinecite{Raikh,Fogler_chaos}),
\begin{eqnarray}
\tilde{U}_0(\bbox{q}) &=& \tilde{U}(\bbox{q})\, F_N(q l^2),
\label{U_0}\\
\displaystyle F_N(y) &=& L_N\left(\frac{y^2}{2 l^2}\right)\,
e^{-y^2 / 4 l^2},
\label{F_N}
\end{eqnarray}
$L_N(z)$ being the Laguerre polynomial (the tilde over the symbol itself
indicates the two-dimensional Fourier transform). 

As discussed in the previous section, in the absence of the random
potential, $\rho_y$ is a good quantum number. The tunneling requires
propagation under the magnetic barrier, which leads to a
super-exponential decay of $G(0, y; E)$. Indeed, in the absence of the
random potential $G_\rho(\rho_1, \rho_2; E) = \delta(\rho_1 - \rho_2) /
E$ [see Eq.~(\ref{Dyson})]. Upon substitution into
Eq.~(\ref{transform_Green}) one recovers the well-known expression for
the Green's function in the clean case,
\begin{equation}
\displaystyle G(0, y; E) = \frac{F_N\left(y\right)}{2\pi l^2 E}.
\label{Green_0}
\end{equation}
Since $F_N(y)$ is the product of a polynomial and a Gaussian [see
Eq.~(\ref{F_N})], the Green's function $G(0, y; E)$ decays faster than the
exponential at large $y$.  In view of definition~(\ref{decay_Green}) this
means that $\xi = 0$.  In fact, the structure of Eq.~(\ref{transform_Green})
suggests that nonzero $\xi$, i.e., a simple exponential decay of $G$, is
possible only if $G_\rho$ decays no faster than a simple exponential.  In
other words, $\xi$ is nonzero only if $\xi_\rho$ is nonzero where
\begin{equation}
\frac{1}{\xi_\rho} =  -\lim\limits_{\rho\to\infty}\frac{1}{\rho}
          \left\langle \ln|G_\rho(0, \rho; E)| \right\rangle
\label{xi_rho}
\end{equation}
[compare with the definition of $\xi$, Eq.~(\ref{decay_Green})].
Unfortunately, $\xi$ and $\xi_\rho$ may differ. Only
the inequality
\begin{equation}
                            \xi_\rho \geq \xi
\label{xi_and_xi_rho}
\end{equation}
%

%
%
\begin{figure}
\centerline{
\psfig{file=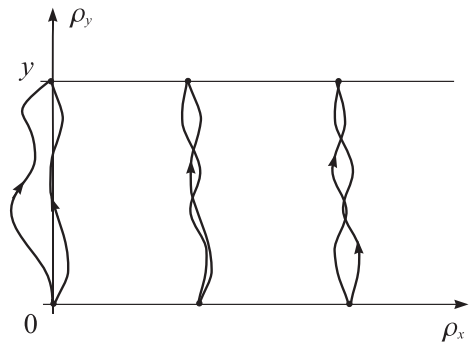,width=2.4in,bbllx=107pt,bblly=247pt,bburx=495pt,bbury=531pt}
}
\vspace{0.1in}
\setlength{\columnwidth}{3.2in}
\centerline{\caption{
Tunneling paths of the guiding center (schematically). Green's function
$G(0, y; E)$ is a sum over the paths near $\rho_x = 0$; Green's function
$G_\rho(0, y; E)$ is a sum over all the possible paths connecting the lines
$\rho_y = 0$ and $\rho_y = y$.
\label{paths}
}}
\end{figure}
\noindent is guaranteed to be met. Indeed, $G_\rho(\rho_1, \rho_2; E)$ typically
behaves like $G_\rho \sim e^{-|\rho_2 - \rho_1| / \xi_\rho +
i\phi(\rho_1, \rho_2)}$. If the phase $\phi(\rho_1, \rho_2)$ is a smooth
function of coordinates, then $|G(0, y; E)| \sim |G_\rho(0, y; E)|$, and
$\xi = \xi_\rho$. Otherwise, the integrand in
Eq.~(\ref{transform_Green}) oscillates rapidly, $|G(0, y; E)| \ll
|G_\rho(0, y; E)|$ and $\xi < \xi_\rho$. On a physical level, $\xi$
describes the tunneling between two point-like contacts while $\xi_\rho$
characterizes the tunneling between two infinite parallel leads. While
the Feynman paths contributing to the former process make up a narrow
bundle near $\rho_x = 0$ (Fig.~\ref{paths}), the latter one gathers
contributions of many such bundles. As a result, the amplitude of the
latter process is much larger on the account of rare places
(``pinholes'') where the tunneling is unusually strong.

Nevertheless, in many cases $\xi$ and $\xi_\rho$ are very close to each
other; for instance, when the random potential is short-range (see the
next section).

Next we would like to present a simple yet very instructive model. This
model has a great advantage of being solvable.

Suppose that the averaged random potential has the form
\begin{equation}
    U_0(\bbox{\rho}) = U_1(\rho_y)
                        + e^{i\rho_x q} U_2(\rho_y)
                        + e^{-i\rho_x q}\, U_2^\ast(\rho_y),
\label{U_0_model}
\end{equation}
where $U_1$, ${\rm Re}\,U_2$, and ${\rm Im}\,U_2$
are mutually independent Gaussian random variables. Similar to the above,
we will assume that they have amplitude $W$ and correlation length $d$.
A simpler model with $\rho_y$-independent $U_1$, ${\rm Re}\,U_2$, and ${\rm Im}\,U_2$
has been studied in Refs.~\onlinecite{Raikh_xi} and~\onlinecite{Haidu}.

For the model potential~(\ref{U_0_model}) all the points along the
$\rho_x$ axis are statistically equivalent; the ``pinholes'' we
mentioned above are absent; therefore, $\xi = \xi_\rho$.

\begin{figure}
\centerline{
\psfig{file=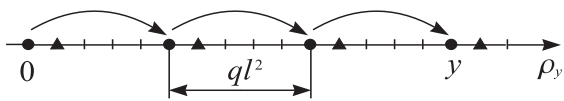,width=2.9in,bbllx=126pt,bblly=395pt,bburx=481pt,bbury=455pt}
}
\vspace{0.1in}
\setlength{\columnwidth}{3.2in}
\centerline{\caption{
The model system. Each vertical tick corresponds to one of the basis
states $\psi_n$. The distance between the ticks is $2\pi l^2 / L_x$. The
hopping (symbolized by arrows) is possible at (much larger) distance $q
l^2$, and takes place between the states forming an equidistant chain.
Two of such chains (one marked by the dots and the other by the
triangles) are shown.
\label{chains}
}}
\end{figure}

From Eq.~(\ref{Dyson}), we see that the matrix element $\langle
\rho_1 | U_0 | \rho_2 \rangle$ is zero unless $\rho_1 = \rho_2$ or
$\rho_1 - \rho_2 = \pm q l^2$. It is convenient to assume that $q l^2$
is divisible into $2 \pi l^2 / L_x$, the smallest distance between the
centers of gravity of the basis states~(\ref{basis}). In other words, we
will assume that $L_x q / (2 \pi)$ is an integer. Under this condition,
the system can be split into $L_x q / (2 \pi)$ independent chains
(Fig.~\ref{chains}). The guiding center coordinates $\{\rho_n\}$ in each
chain form an equidistant set: $\rho_{n + 1} - \rho_n = q l^2$. The
hopping is allowed only between the nearest neighbors of the same chain
and is characterized by the hopping amplitude $U_2^\ast[(\rho_n +
\rho_{n + 1}) / 2]$. As for $U_1(\rho_n)$, it plays the role of the
on-site energy.

%
%

The localization length of a disordered chain is given
by the exact formula due to Thouless,~\cite{Thouless} which in our case
reads
\begin{equation}
\frac{q l^2}{\xi} = \int\! d E' D(E') \ln|E - E'| 
- \left\langle \ln|U_2(\rho)| \right\rangle,    
\label{Thouless_xi}
\end{equation}
$D(E)$ being the disorder-averaged density of states normalized by the
condition $\int\! d E\, D(E) = 1$. It is noteworthy that $D(E)$ can in
principle be found exactly if all the matrix elements are statistically
independent,~\cite{Dyson} i.e., if $q l^2 \gg d$. If this is not the
case, then $D(E)$ can be calculated by some approximation scheme. At any
rate, $D(E)$ is small if $|E| \gg W$ (see e.g., Lifshitz {\it et
al.\/}~\cite{Lifshitz}). In this case we can expand the logarithm in
Eq.~(\ref{Thouless_xi}) in the powers of $E'/ E$ to obtain
\begin{equation}
\frac{1}{\xi} = \frac{1}{q l^2}\left[
               \left\langle\!\ln\left|\frac{E}{U_2}\right|\,\right\rangle
               + O\left(\frac{W}{E}\right)^2 \right].
\label{Thouless_xi II}
\end{equation}
Taking the average in Eq.~(\ref{Thouless_xi II}), we obtain
\begin{equation}
\frac{1}{\xi} = \frac{1}{q l^2}
               \left[\ln\left|\frac{E}{W}\right| - \frac{\ln 2 - {\bf C}}{2}
               + O\left(\frac{W}{E}\right)^2 \right],
\label{xi_chain}
\end{equation}
where ${\bf C} = 0.577\ldots$ is the Euler constant.

It is quite remarkable that $\xi$ does not depend on whether the
successive hopping terms $U_2^\ast[(\rho_n + \rho_{n + 1}) / 2]$ are
correlated or not. With the high degree of accuracy, $O(W^2 / E^2)$, the
localization length has the same value for $q l^2 > d$ (``short-range''
disorder) and $q l^2 < d$ (``long-range'' disorder).

The qualitative derivation of Eq.~(\ref{xi_chain}) can be done with the
help of the locator expansion (see a similar argument in the preceding
section). The tunneling through a distance $y$ is achieved by a minimum
of $M = y / (q l^2)$ hops. Each hop is characterized by the hopping
amplitude of the order of $W / E$. Thus, the suppression factor of the
wave function over a distance $y$ is of the order of $(W / E)^M$. On the
other hand, this factor is equal to $e^{-y / \xi}$, which leads to
Eq.~(\ref{xi_chain}).

Let us now return to the original problem with the two-demensional
random potential [Eqs.~(\ref{C_r}) and~(\ref{C_q})]. Leaving a more rigorous
calculation for the next two sections, we will present heuristic
arguments leading to Eqs.~(\ref{xi_short}) and~(\ref{xi_white}).

Let us divide the entire spectrum of Fourier harmonics of the random
potential $U_0$ into bands $q_n - \frac12\Delta q < q_x < q_n +
\frac12\Delta q$, $n = 1, 2,\ldots$ of width $\Delta q \sim 1 / d$.
Denote by $U_q(\bbox{\rho})$ the combined amplitude of the harmonics,
which make up the band centered at $q$,
\begin{equation}
U_q(\bbox{\rho}) = \int\limits_{q - \frac12\Delta q}^{q + \frac12\Delta q}
\!\!\frac{d q_x}{2 \pi}\, {U}_0(\tilde{q}_x, \rho_y)\, e^{i \rho_x q_x},
\label{U_q}
\end{equation}
If $q \gg 1 / d$, the corresponding band is very narrow, and $U_q$
as a function of $\rho_x$ looks very much like a plain wave,
$U_q \propto e^{i q \rho_x}$, exactly as $U_2$ in the model problem.
Suppose that the scattering caused by different bands
can be considered independently. In this case each band generates
its own decay rate $1 / \xi(q)$ of the wave function. By analogy with
Eq.~(\ref{Thouless_xi II}), we can write
\begin{equation}
\frac{1}{\xi(q)} \simeq \frac{1}{2 q l^2}
\left\langle\!\ln\left|\frac{E^2}{W_q^2}\right|\,\right\rangle,
\label{xi_band}
\end{equation}
where $W_q^2$ is the variance of $U_q(\bbox{\rho})$,
\begin{equation}
            W_q^2 \sim C_0(\tilde{q}, 0)\, \Delta q,
\label{W_q}
\end{equation}
and $\tilde{C}_0(q)$ is the correlator of the averaged potential,
\begin{equation}
           \tilde{C}_0(q) = \tilde{C}(q) \left[F_N(q l^2)\right]^2.
\label{C_0}
\end{equation}
Let $q_\ast$ be the wave vector corresponding to the largest $\xi(q)$,
then it is natural to think that $\xi = \xi(q_\ast)$. In other words, the
localization length should be determined by the ``optimal band'' of
harmonics, which we are going to find next.

In view of Eq.~(\ref{C_0}), two cases have to be distinguished,
$q_\ast < 2 k_F$ and $q_\ast > 2 k_F$. The latter
is realized for a sufficiently weak ``white-noise'' random potential,
the former for the potentials of all other types.

If $q < 2 k_F$, then $C_0(\tilde{q}, 0)$ differs from $C(\tilde{q}, 0)$
only by a pre-exponential factor. Using Eqs.~(\ref{C_q})
and~(\ref{W_q}) and omitting some unimportant pre-exponential
factors, we arrive at
\begin{equation}
\frac{1}{\xi(q)} \simeq \frac{1}{q l^2}\left[
   \ln\left|\frac{E}{W}\right| + \frac{(q d)^\beta}{2\beta}\right].
\label{xi_q}
\end{equation}
If $\beta > 1$, then $\xi(q)$ given by Eq.~(\ref{xi_q}) has the
maximum at
\begin{equation}
q_\ast(E) = \frac{1}{d} \left(\frac{2\beta}{\beta - 1}
\ln\left|\frac{E}{W}\right|\right)^{\frac{1}{\beta}}.
\label{q_0}
\end{equation}
Substituting this value into Eq.~(\ref{xi_q}) and taking $E =
\hbar\omega_c / 2$, we obtain Eq.~(\ref{xi_short}).

The qualitative derivation of Eq.~(\ref{xi_white}) goes along the same
lines. The sole difference is that $q_\ast$ turns out to be close or
even larger that $2 k_F$ and at the same time smaller than $1 / d$. In
this case $\tilde{C}_0(q_\ast)$ is determined by $F_N(q_\ast l^2)$
rather than by $\tilde{C}(q_\ast)$. (In this case, of course, the
appropriate width $\Delta q$ of the bands is much smaller than $1 / d$,
but the basic idea of dividing the spectrum into independent bands
stays).

\section{Short-range random potential}
\label{short}

In this section we present a more detailed calculation of the
localization length for the short-range random potential, $d \ll l$. As
we mentioned in the preceding section, Thouless
formula~(\ref{Thouless_xi II}) is in agreement with the locator
expansion for $G_\rho$,
\begin{equation}
G_\rho(0, M q l^2; E) \simeq \frac{L_x}{2 \pi l^2 E} \prod_{m = 1}^{M}
\frac{U_2^\ast\left[\left(m - \frac12\right) q l^2\right]}{E}.   
\label{locator_chain}
\end{equation}
The equivalent of Eq.~(\ref{locator_chain}) in the general case is
\begin{eqnarray}
&\displaystyle G_\rho(0, y; E) \simeq \frac{1}{E}\sum_{M = 0}^\infty
\int \frac{d q_{1}}{2 \pi}
\int \frac{d q_{2}}{2 \pi} \ldots
\int \frac{d q_{M}}{2 \pi}&\nonumber\\
&\displaystyle \times \delta\left(y - l^2 \sum_{n = 1}^{M} q_{n}\right)
\prod_{m = 1}^{M}
\frac{U_0^\ast\left(\tilde{q}_{m}, \rho_{m y}\right)}{E},&   
\label{locator}
\end{eqnarray}
where $\rho_{m y}$ stands for
\begin{equation}
\rho_{m y} = l^2 (q_{1} + q_{2} +\ldots + q_{m - 1} +
\frac12 q_{m}).
\label{rho_my}
\end{equation}
Combining Eqs.~(\ref{transform_Green}) and~(\ref{locator}), we arrive at
\begin{eqnarray}
\displaystyle & & G^{(M)}(0, y; E) = \sum_{M = 0}^\infty G^{(M)},
\\
\displaystyle & & G^{(M)} \simeq \int \frac{d \rho}{2 \pi l^2 E}\,
\Phi_N(\rho)\, \Phi_N(\rho + \Sigma - y)
\nonumber\\
\displaystyle & & \times \int\ldots\int\! \prod_{m = 1}^M
\frac{d q_{m}}{2 \pi E}\, U_0^\ast\left(\tilde{q}_{m}, \rho + \rho_{m y}\right),   
\label{Green_M}
\end{eqnarray}
where $\Sigma = \sum_{n = 1}^M q_{n} l^2$.

Formula~(\ref{locator}) is certainly just an approximation. However, the
model studied in the preceding section showed that the the relative
error in calculating $\xi$ in this way is of the order $O\left(W /
E\right)^2$, which is quite satisfactory. The major defect of our
approximation is having all the energy denominators equal to $E$.
Consequently, this approximation scheme does not capture the phenomenon
of the resonant tunneling,~\cite{Lifshitz} which appears due to
anomalously small energy denominators. Note however that our goal is to
calculate $\langle \ln |G(0, y; E)| \rangle$. The resonant tunneling
configurations are exponentially rare and do not contribute to this
quantity. In the model studied in the previous section this can be
seen explicitly: the resonant tunneling configurations correspond to
$E = E'$ in Eq.~(\ref{Thouless_xi}) where the integrand diverges.
However, the divergence is integrable, and moreover has an exponentially small
weight.

We calculate $\langle \ln |G(0, y; E)| \rangle$ in three steps. First,
we calculate $\langle |G(0, y; E)|^2 \rangle_{\rm nr}$, where the
subscript ``nr'' stands for ``non-resonant,'' i.e., with resonant
tunneling configurations excluded. The reminder of such an exclusion is
essential in this case because even being exponentially rare, the
resonant tunneling configurations yield untypically large $|G(0, y; E)|^2$
and totally dominate the average square modulus $\langle |G(0, y; E)|^2
\rangle$ for sufficiently large $y$ (see Ref.~\onlinecite{Shklovskii_83}
and Ref.~\onlinecite{Comment on Ward}).

As the next step, we calculate the decay rate of
$\langle |G(0, y; E)|^2 \rangle_{\rm nr}$ defined similarly
to Eq.~(\ref{decay_Green}),
\begin{equation}
\frac{1}{\xi_2} = -\lim\limits_{y\to\infty}\frac{1}{2 y}
           \ln\left\langle|G(0, y; E)|^2 \right\rangle_{\rm nr}.
\label{decay_G_squared}
\end{equation}

Finally, we show that $\xi_2 = \xi$.

\begin{figure}
\centerline{
\psfig{file=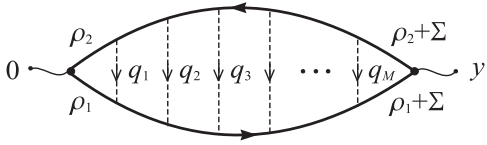,width=2.5in,bbllx=128pt,bblly=383pt,bburx=466pt,bbury=484pt}
}
\vspace{0.1in}
\setlength{\columnwidth}{3.2in}
\centerline{\caption{
A typical ladder diagram, which describes the tunneling in the
short-range random potential.
\label{ladder}
}}
\end{figure}

The calculation of $\langle |G(0, y; E)|^2 \rangle_{\rm nr}$ can by
represented with the help of diagrams, one of which is shown in
Fig.~\ref{ladder}. The solid lines in this diagram correspond to the
factors $1 / E$, each dashed line stands for $C_0(\tilde{q}, \rho)$ with
appropriate arguments, and the vortices at the corners bring the factors
$\Phi_N(\cdot) \Phi_N(\cdot)$.

%
%

The diagram shown in Fig.~\ref{ladder} is of the ladder type. It is easy
to see that other diagrams (with crossing dashed lines) are negligible.
Indeed, consider, for instance, a diagram where $m$th and $m + 1$st dashed
lines of the original ladder diagram are interchanged. This diagram will
be proportional to $C_0(\tilde{q}_{m}, \Delta\rho_y)$ where
$\Delta\rho_y = \rho_{m, y} - \rho_{m + 1, y} \sim q_{m} l^2$. As
discussed in the preceding section, the characteristic values of $q_{m}$
are of the order of $q_\ast(E)$, so that the distance $\Delta\rho_y$
between $\rho_{m y}$ and $\rho_{m + 1, y}$ is of the order of
$q_\ast(E)\, l^2$. This distance is much larger than $d$, the
correlation length of $U_0$ because $l \gg d$. Therefore,
$C_0(\tilde{q}_{m}, \Delta\rho_y)$, and thus, the entire diagram are
small.

Note also that the neglect of the resonant tunneling configurations is
assured by omitting the diagrams with dashed lines connecting two points
of the same solid line (either the upper or the lower one).

The magnitude of the ladder diagram in Fig.~\ref{ladder} is equal to
\wide{m}{
\begin{equation}
\langle |G^{(M)}|^2 \rangle_{\rm nr} = \int\!\!\int\!
\frac{d \rho_1 d \rho_2}{(2 \pi l^2 E)^2}
[\Phi_N(\rho_1)\, \Phi_N(\rho_2)]
[\Phi_N(\rho_1 + \Sigma - y)\, \Phi_N(\rho_2 + \Sigma - y)]
\int\!\!\ldots\!\!\int\!\! \prod_{m = 1}^M
\frac{d q_{m}}{2 \pi E^2}\, C_0\left(\tilde{q}_{m}, \rho_2 - \rho_1\right).
\label{G_M_ladder}
\end{equation}
}
The products of functions $\Phi_N$ in the square brackets can be
replaced by the integrals over auxiliary variables according to the
formula
\begin{equation}
\Phi_N(x + \frac{y}{2})\, \Phi_N(x - \frac{y}{2}) = 
\int\! \frac{d z}{2 \pi l^2}\, e^{i z x / l^2}\, F_N(z, y),
\label{Phi_Phi}
\end{equation}
where $F_N(z, y) = F_N(\sqrt{z^2 + y^2})$,
which enables one to obtain a rather simple expression,
\begin{equation}
\langle |G^{(M)}|^2 \rangle_{\rm nr} = \int\!
\frac{d^2 \rho\, F_N^2(\rho)}{(2 \pi l^2)^3}\,
\frac{C_0(\rho)^M\, e^{i \rho_x y / l^2}}{E^{2 M + 2}},
\end{equation}
and finally,
\begin{equation}
\langle |G(0, y; E)|^2 \rangle_{\rm nr} = \int\!
\frac{d^2 \rho\,F_N^2(\rho)}{(2 \pi l^2)^3}\,
\frac{e^{i \rho_x y / l^2}}{E^2 - C_0(\rho)}.
\label{G_squared}
\end{equation}
A very similar calculation gives
\begin{equation}
\langle |G_\rho(0, y; E)|^2 \rangle_{\rm nr} = L_x \int\!
\frac{d \rho}{(2 \pi l^2)^2}\, \frac{e^{i \rho y / l^2}}{E^2 - C_0(\rho)}.
\label{G_rho_squared}
\end{equation}
Since the integrands in these formulas oscillate the more rapidly the
larger $y$ is, the square modulus of the two Green's functions decays
with $y$. Furthermore, comparing Eqs.~(\ref{G_squared}) and~(\ref{G_rho_squared}), we see that
the decay rate of $G$ and $G_\rho$ is the same, i.e., that
$\xi_2$ would not change if $G$ were replaced by $G_\rho$ in
Eq.~(\ref{decay_G_squared}). In both cases $\xi_2$ is given by
\begin{equation}
                     \xi_2 = l^2 / v_\ast,
\label{xi_2}
\end{equation}
where $v_\ast$ is the smallest positive root of the equation
\begin{equation}
C_0(2 i v_\ast, 0) \equiv \int \! \frac{d q_x d q_y}{(2 \pi)^2}\,
C_0(\tilde{q}_x, \tilde{q}_y)\, e^{2 q_x v_\ast} = E^2.
\label{v_0}
\end{equation}
Using Eqs.~(\ref{C_q}) and~(\ref{C_0}), and also the asymptotic formula $[F_N(q
l^2)]^2 \simeq 1 / (\pi q R_c)$ (valid for $N \gg 1$ and $q
\ll k_F$) we obtain
\begin{equation}
\int\limits_0^\infty \!
\frac{d q}{\pi R_c (4 \pi q v_\ast)^{1/2}}\,
e^{-\frac{1}{\beta}(q d)^\beta + 2 q v_\ast}
\simeq \frac{E^2}{\tilde{C}(0)}.
\label{v_0_integral}
\end{equation}
If $\beta > 1$, the integrand has the saddle-point at $q = q_\ast$
[Eq.~(\ref{q_0})] with the characteristic spread of $q$ around $q_\ast$
being $\Delta q \sim 1 / \{d\,[\ln(E / W)]^{(\beta - 2) / 2 \beta}\}$.
Using the saddle-point method estimate for the integral, and then
solving the resulting transcendental equation, we obtain
\begin{equation}
v_\ast \simeq \left(\frac{2\beta}{\beta - 1} \ln \frac{|E|}{W}
           \right)^{\frac{\beta - 1}{\beta}} \frac{d}{2}.
\label{v_0_saddle}
\end{equation}
As one can see from Eqs.~(\ref{xi_2}) and~(\ref{v_0_saddle}), the
derivation of Eq.~(\ref{xi_short}) will be complete if we demonstrate
that $\xi_2 = \xi$, i.e., that $\ln \langle |G(0, y; E)|^2 \rangle_{\rm
nr} - \langle \ln |G(0, y; E)|^2 \rangle = o(y)$. (Note that we are
interested mainly in the case $|E| = \hbar\omega_c / 2$). Since such a
calculation is not an easy task, we will only show that this relation
holds for $G^{(M)}$, where $M \approx y / (q_\ast l^2)$. Such $M$ give
dominant contribution to $\langle |G(0, y; E)|^2 \rangle_{\rm nr}$, and
presumably, to $\langle \ln |G(0, y; E)|^2 \rangle$ as well.

To average the logarithm, we employ the replica trick,
\begin{equation}
\langle \ln |G| \rangle = \lim_{n \to 0}
\frac{\langle |G|^{2 n} \rangle - 1}{2 n}.
\label{replica_trick}
\end{equation}
Under the same kind of approximations as above
and for integer $n$, $\langle |G^{(M)}|^{2 n} \rangle_{\rm nr}$ is given by
\wide{t}{
\begin{equation}
\langle |G^{(M)}|^{2 n} \rangle_{\rm nr} = \int\ldots\int\!
\prod_{m = 1}^{2 n} \frac{d \rho_m}{2 \pi l^2 E}
\Phi_N(\rho_m)\,\Phi_N(\rho_m + \Sigma_m - y)
\int\ldots\int\! \prod_{k = 1}^M \prod_{r = 1}^n
\frac{d q_{k}^{(r)}}{2 \pi E^2}
\sum_{P_k}
C_0\left[\tilde{q}_{k}^{(r)}, \rho_{k y}^{(r)} - \sigma_{k y}^{(r)}\right],
\label{G_M_2n}
\end{equation}
}
where $\Sigma_m = \sum_{k = 1}^M q_{k}^{(m)} l^2$ and $P_k$ labels the
permutations of the superscripts of the set $\{q_k^{(1)},
q_k^{(2)}, \ldots, q_k^{(n)}\}$. The quantity $\sigma_{k y}^{(r)}$
stands for $l^2 [q_1^{P_1(r)} + \ldots + q_{k - 1}^{P_{k - 1}(r)} +
\frac12 q_{k}^{P_{k}(r)}]$. There are altogether $n!$ permutations
$P_k$ for each $k$; therefore, the complete expression is a rather
complicated sum of $(n!)^M$ terms. However, only $n!$ terms in this sum
are significant. Indeed, within the adopted approximation all the terms
with $|\rho_{k y}^{(r)} - \sigma_{k y}^{(r)}| \gg d$ for at least one of
$k = 1,2,\ldots, M$, should be dropped. It is easy to find the
difference $\rho_{k y}^{(r)} - \sigma_{k y}^{(r)}$ for the case of
identical permutations $P_1, P_2,\ldots,P_M$. In this case $\rho_{k
y}^{(r)} - \sigma_{k y}^{(r)} = \rho_r - \rho_{P_1(r)}$ for all $k$.
Therefore, a single constraint $|\rho_r - \rho_{P_1(r)}| \lesssim d$
takes care of all the $M$ constraints above. If, on the other hand, some
of $P_k$ are not the same, then the integration domain is much more
restricted, and the value of the integral is small. Retaining only the
terms corresponding to identical permutations, we immediately find that
\begin{equation}
\langle |G^{(M)}|^{2 n} \rangle_{\rm nr} \simeq n! \,
\langle |G^{(M)}|^2 \rangle^n_{\rm nr}.
\end{equation}
Taking the limit $n \to 0$, we obtain from here~\cite{Comment on Shklovskii_83}
\begin{equation}
\langle \ln |G^{(M)}| \rangle \simeq
\frac{\ln \langle |G^{(M)}|^2 \rangle_{\rm nr} - {\bf C}}{2}.
\end{equation}
We think that the same relation holds if $G^{(M)}$ is replaced by
the total Green's function, $G(0, y; E)$, and so $\xi_2 = \xi$.

Finally, let us sketch the derivation of $\xi$ for the ``white-noise''
random potential, $d \ll k_F^{-1}$. In this limit one can replace
$\tilde{C}_0(\bbox{q})$ by $\tilde{C}(0) F_N^2(q l^2)$ in Eq.~(\ref{v_0}),
which leads to the following equation on $v_\ast$,
\begin{equation}
         \frac{\tilde{C}(0)}{2 \pi l^2}\, F_N^2(2 i v_\ast) = E^2.
\label{v_0_white}
\end{equation}
The next step is to use the asymptotic formula for $F_N(i y)$,
\begin{eqnarray}
\displaystyle F_N(i y) &\simeq&
\frac{l}{\sqrt{\pi y s}}
\left(\frac{y + s}{2 R_c}\right)^{2 N + 1} e^{{y s}/{4 l^2}},
\label{F_N_asym}\\
s &=& \sqrt{y^2 + 4 R_c^2},
\end{eqnarray}
valid for $y \gg k_F^{-1}$. This way one obtains an approximate solution
for $v_\ast$. Finally, taking $|E|$ to be $\hbar\omega_c / 2$, 
one recovers Eq.~(\ref{xi_white}).

\section{Long-range random potential: discrete Landau levels}
\label{long}

In the previous section devoted to short-range random potentials,
we were able to derive $\xi$ by calculating the square modulus of
the Green's function. Unfortunately, this is not possible for
a long-range random potential. The physical reason is as follows.

As we have shown in Sec.~\ref{single}, the typical distance
$\Delta\rho_y$ between the locations of successive scattering events is
of the order of $(l^2 / d) {\cal L}$, where ${\cal L}$ is some
logarithmic factor. If the random potential is long-range, $d \gg l$,
then $\Delta\rho_y \ll d$, and so such scattering events can no longer
be considered uncorrelated. One of consequences is an enhanced
probability of ``pinholes'' discussed in Sec.~\ref{single}. In other
words, the local decay rate of the wave functions with distance becomes
very nonuniform. In its turn, the Green's function $G(0, y; E)$, even
with the resonant tunneling configurations excluded, exhibits
large fluctuations between different disorder realizations so that $|\ln
\langle |G(0, y; E)|^2 \rangle| \gg |\langle \ln |G(0, y; E)|^2
\rangle_{\rm nr}|$, or $\xi_2 \gg \xi$.

One look at Eq.~(\ref{G_M_2n}) is sufficient to predict that a
diagrammatic calculation of $\langle \ln |G(0, y; E)|^2\rangle$ is bound
to be very cumbersome. The task is easier within a
different approximation scheme, the WKB method. Some of the formulas
corresponding to this approximation have been previously worked out by
Tsukada~\cite{Tsukada} and by Mil'nikov and Sokolov.~\cite{Milnikov}

Suppose we want to find the solution of the Schr\"oedinger
equation $\hat{U}_0 \phi(\rho_y) = E \phi(\rho_y)$. Let us seek the
solution in the form $\phi(\rho_y) = \exp[i S(\rho_y)]$. The action
$S(\rho_y)$ can be expanded in the series of the small parameter $l / d$.
In the lowest approximation, $S(\rho_y)$ must satisfy the Hamilton-Jacoby
equation,
\[
   U_0\left(l^2 \frac{\partial S}{\partial\rho_y}, \rho_y\right) = E,
\]
so that
\[
\phi(\rho_y) \sim \exp\left[\frac{i}{l^2}\! \int\limits^{\rho_y}\! d \eta\,
                            \rho_x(\eta) \right],
\]
where $\rho_x(\rho_y)$ is a solution (generally speaking, a complex one) of the
equation
\begin{equation}
                      U_0(\rho_x, \rho_y) = E.
\label{level_line}
\end{equation}
The meaning of this equation is quite transparent. It is known that the
motion of the guiding center in classically permitted regions is a drift
along the level lines of the averaged potential $U_0$ (see
Ref.\onlinecite{Fogler_chaos}). Equation~(\ref{level_line}) means that
the trajectory of the guiding center in classically forbidden regions is
still a level line although obtained by analytical continuation to the
three-dimensional space $\{(u, v, \rho_y)\}$ where $u = {\rm Re}\, \rho_x$
and $v = -{\rm Im}\, \rho_x$.

The WKB-type formula for the Green's function in the guiding center
representation is,
\wide{t}{
\begin{equation}
G_\rho(\rho_1, \rho_2; E) = \frac{i}{l^2} \sum_n
\frac{{\rm sgn}(\rho_2 - \rho_1)}{\sqrt{
\displaystyle \left[
{\partial U_0}/{\partial\rho_x^{(n)}}\right]_{\rho_y = \rho_1}
              \left[
{\partial U_0}/{\partial\rho_x^{(n)}}\right]_{\rho_y = \rho_2}}}\,
\exp\left[-\frac{i}{l^2} \int\limits_{\rho_1}^{\rho_2}\!
d \rho_y\,\, \rho_x^{(n)}(\rho_y) \right],
\label{G_rho_WKB}
\end{equation}
}
where the superscript $n$ labels different solutions $\rho_x^{(n)} = u_n
- i v_n$ of Eq.~(\ref{level_line}) in the complex half-space ${\rm
sgn}\,v_n = {\rm sgn}(\rho_2 - \rho_1)$. If we study the tunneling from
point $(0, 0)$ to $(0, y)$ where $y > 0$, then this will typically be the
upper half-space, $v_n \geq 0$. Using Eqs.~(\ref{transform_Green})
and~(\ref{G_rho_WKB}) and neglecting all the pre-exponential factors, we
obtain the following estimate for $G(0, y; E)$,
\[
G \sim \sum_n \exp\!\!\left[
-\frac{u_n(0)^2 + u_n^2(y)}{2 l^2}
-\frac{i}{l^2}\! \int\limits_{0}^{y}\!
d \rho_y\,\, \rho_x^{(n)}(\rho_y) \right].
\]
Since $U_0$ is a random potential, it is natural to assume that the
phase factors corresponding to different $n$ in this sum are
uncorrelated; therefore,
\begin{equation}
|G|^2 \sim \!\sum_n \exp\!\!\left[ -\frac{u_n(0)^2 + u_n^2(y)}{l^2}
-\frac{2}{l^2}\!\! \int\limits_{0}^{y}\! d \rho_y\, v_n(\rho_y) \right].
\label{G_WKB II}
\end{equation}
Consequently, $\xi$ can be calculated as follows,
\begin{equation}
    \frac{1}{\xi} = \min \limits_n \lim_{y \to \infty} \frac{1}{y l^2}
\left[\frac{u_n^2(y)}{2} + \int\limits_0^y\! d\rho_y\,\, v_n(\rho_y)\right].
\label{one_xi}
\end{equation}
It is possible to demonstrate that the first term in the square brackets
is typically much smaller than the second one, which leads to
\begin{equation}
            \xi \simeq \frac{l^2}{\min\limits_n\, \langle v_n \rangle},
\label{xi_v}
\end{equation}
where $\langle v_n \rangle$ is the average ``height'' of the $n$th
level line,
\[
\langle v_n \rangle = \lim_{y \to \infty} \frac{1}{y}
\int\limits_0^y\! d\rho_y\,\, v_n(\rho_y).
\]

The problem of calculating $\min\limits_n\, \langle v_n \rangle$ turns
to be rather difficult and we have not been able to solve it exactly.
However, we will give arguments that $\min\limits_n\, \langle v_n
\rangle \simeq v_\ast$ [$v_\ast$ was introduced in the previous section,
see Eq.~(\ref{v_0})]. Therefore, $\xi$ is still given by
formula~(\ref{xi_short}). A more accurate statement is as follows. If
the individual Landau levels are well resolved in the density of states,
then for arbitrary range $d$ of the random potential 
$\xi$ can be found from the same ``master'' equation
\begin{equation}
                     C_0(2 i l^2 / \xi, 0) = E^2.
\label{xi_master}
\end{equation}

Let us familiarize ourselves with the properties of the equipotential
contours (level lines) of the potential $U_0$ in the half-space $\{(u,
v, \rho_y), v \geq 0\}$. An important property
is the contour density $P(v, E)$,
\[
 P(v, E) = \left\langle \sum_n \delta(v - v_n)\,\delta(u - u_n) \right\rangle
\]
(for isotropic random potential $U_0$ this quantity does not depend neither
on $u$ nor on $\rho_y$). Function $P(v, E)$ proves to be the sum of three
terms,
\begin{equation}
          P(v, E) = P_1(v, E) + P_2(v, E) + P_3(E)\, \delta(v),
\label{P}
\end{equation}
where
\begin{eqnarray}
\displaystyle P_1(v, E) &=& \frac{E^2 S^2}{\pi Q^2 \sqrt{Q R}}\,e^{-E^2 / Q},
\label{P_1}\\
\displaystyle P_2(v, E) &=& \frac{e^{-E^2 / Q}}{2\pi \sqrt{Q R}}
\left[T - S^2 \left(\frac{1}{Q} + \frac{1}{R}\right)\right],
\label{P_2}
\end{eqnarray}
and $Q$, $R$, $S$, and $T$ are as follows,
\begin{equation}
\begin{array}{c}
Q = C_0(2 i v, 0) + C_0(0, 0),\\[12pt]
R = C_0(2 i v, 0) - C_0(0, 0),\\[12pt]
S = \frac12\, {d Q}/{d v},\:\:\: T = \frac12\,{d^2 Q}/{d v^2}.
\end{array}
\label{QRST}
\end{equation}
Except for the term $P_3(v)$ our formulas are in agreement with
Refs.~\onlinecite{Raikh_xi} and~\onlinecite{Shahbazyan}. This term,
however, does not play any role in the subsequent calculation. The
derivation of Eqs.~(\ref{P_1}) and~(\ref{P_2}) can be found in the
Appendix.

%
%
\begin{figure}
\centerline{
\psfig{file=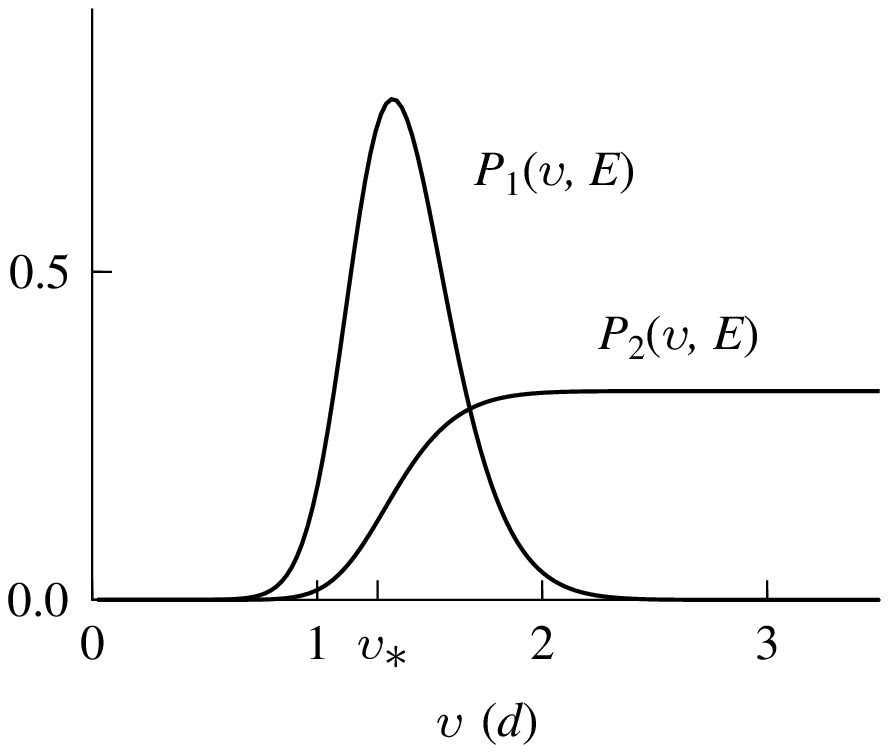,width=2.7in,bbllx=194pt,bblly=494pt,bburx=442pt,bbury=710pt}
}
\vspace{0.1in}
\setlength{\columnwidth}{3.2in}
\centerline{\caption{
Functions $P_1(v, E)$ and $P_2(v, E)$. Vertical axis is in units of $1 / d^2$.
The parameters used in generating the plot are $\beta = 2$ and $E / W = 5$.
\label{Plot_P}
}}
\end{figure}

The behavior of functions $P_1$ and $P_2$ is illustrated by
Fig.~\ref{Plot_P}. Function $P_1$ has a sharp maximum at $v = v_\ast$
where $Q \simeq E^2$. Away from the maximum it is exponentially small.
Function $P_2$ is exponentially small at $v < v_\ast$, and assumes the
asymptotic form $P_2(v, E) \propto v^{(2 - \beta) / (\beta - 1)}$ at $v
> v_\ast$. For example, if $\beta = 2$, then $P_2(v, E) \to 1 / (\pi
d^2)$ (see Fig.~\ref{Plot_P}). The ratio $P_1 / P_2$ evaluated at the
point $v_\ast$ is of the order of $\ln|E / W| \gg 1$ in the case of
interest. At $v$ close to $v_\ast$ our expression essentially coincides
with Eq.~(5.20) of Ref.~\onlinecite{Raikh_xi}. Note however, that the
latter equation is off by $4\pi$. 

Let us clarify the origin of the sharp maximum in $P(v, E)$ at $v =
v_\ast$. To this end the concept of bands of harmonics introduced in
Sec.~\ref{single} is very helpful. So, let us consider a band with $q_x$
in the interval $(q - \frac12\Delta q, q + \frac12\Delta q)$ where $q
\gg \Delta q > 0$. The amplitude of each harmonic $U_0(\tilde{q}_x,
\rho_y)$ is enhanced by the factor $e^{q_x v}$ upon the analytic
continuation into the upper half-space. Therefore, the typical value of
the combined amplitude $U_q(\bbox{\rho})$ of the band [see
Eq.~(\ref{U_q}) for definition] grows from $W_q(0) \sim [C_0(\tilde{q},
0) \Delta q]^{1/2}$ at $v = 0$ to $W_q(v) \sim [C_0(\tilde{q}, 0) \Delta
q]^{1/2}\, e^{q v}$ at $v > 0$. Being the product of the rapidly
decreasing function $C_0(\tilde{q}, 0)$ and the exponentially growing
factor $e^{q v}$, this quantity has a sharp maximum at $q = q_\ast(v)$,
\begin{equation}
q_\ast(v) = \frac{1}{d}\left(\frac{2 v}{d}\right)^{\frac{1}{\beta - 1}},
\label{q_v}
\end{equation}
provided that $\beta > 1$ and $v \gg d$. [In view of
Eq.~(\ref{v_0_saddle}), this formula is just another parametrization of
$q_\ast$ originally defined by Eq.~(\ref{q_0}) as a function of $E$].
Generally speaking, the width of the maximum depends on $v$ and
$\beta$ but in a particular example, $\beta = 2$, it is simply $\Delta q
\sim \frac{1}{d}$. Hence, at a given ``height'' $v$ in our
three-dimensional space, the potential $U_0(\rho_x, \rho_y)$ is
typically dominated by the band of harmonics of width $\Delta q$
centered at $q_\ast(v)$. Consequently, the spatial dependence of $U_0$
is almost plain-wave-like, $U_0(\rho_x, \rho_y) \propto e^{i \rho_x
q_\ast(v)}$. This prompts the decomposition
\begin{equation}
         U_0(\bbox{\rho}) = e^{i \rho_x q_\ast(v)} V(\bbox{\rho})
\label{V}
\end{equation}
of $U_0$ into the ``oscillating part'' $e^{i \rho_x q_\ast(v)}$ and a
``smooth part'' $V(\bbox{\rho})$.

The intersection points of the level lines $U_0 = E$ with a vertical
plane $\rho_y = \eta$ satisfy the system of equations
\begin{eqnarray}
v_n &=& -\frac{1}{q_\ast(v_n)} \ln \frac{|V|}{E},
\label{v_n}\\
u_n &=& \frac{1}{q_\ast(v_n)}\arg V + \frac{2 \pi n}{q_\ast(v_n)}.
\label{u_n}
\end{eqnarray}
The modulus $|V|$ of the complex Gaussian variable $V$ has the
Maxwellian distribution with the characteristic width ${\cal W}$ given by
\begin{equation}
{\cal W}^2 \equiv \langle |V(\bbox{\rho})|^2 \rangle \simeq
e^{-2 v q_\ast(v)} Q,
\label{V_squared}
\end{equation}
where $Q$ is defined by Eq.~(\ref{QRST}). In the first approximation let
us neglect the dependence of $V$ on $\rho_x$, (which corresponds to the
limit of infinitesimally narrow band, $\Delta q \to 0$), then $P(v, E)$
is given by
\[
P(v, E) = \frac{q_\ast}{2\pi}\!
\int\limits_0^\infty \frac{2 |V|\,d |V|}{{\cal W}^2}\,
e^{-|V|^2 / {\cal W}^2}
\delta\!\left(\!v + \frac{1}{q_\ast}\ln\frac{|V|}{\cal W}\right),
\]

Doing the integration, we arrive at
\begin{equation}
P(v, E) = \frac{E^2 q_\ast^2(v)}{\pi Q} e^{-E^2 / Q},
\label{P_approx}
\end{equation}
which, in fact, coincides with $P_1(v, E)$ provided that $v \gg d$. In
this approximation $u_n$ form an equidistant set, $u_{n + 1} - u_n \simeq
2\pi / q_\ast \ll 1 / d$, while $v_n$ does not depend on $n$. The
equipotential contours resemble a number of uniformly spaced parallel rods,
which ``soar'' above the real plane $v = 0$, staying very close to the
``standard height'' $v = v_\ast$ for most values of $\rho_y$. Indeed, as
discussed above (see also Fig.~\ref{Plot_P}), function $P_1$ has a sharp
maximum at $v = v_\ast$. Since $d \ln Q / d v \simeq 2 q_\ast$, the
width of the maximum is of the order of $1 / q_\ast$. This is,
of course, can be seen directly from Eq.~(\ref{v_n}): if $|V|$ has its
typical value, ${\cal W}$, then $v_n = v_\ast$. Fluctuations of
$|V|$ change the logarithm (typically) by a number of the order of
unity; therefore, ordinarily $|v_n - v_\ast| \simeq 1 / q_\ast$.
Significant deviations from the ``standard height'' $v_\ast$
are exponentially rare.

We believe that such a description accurately portrays the behavior of
the {\em relevant\/} equipotential contours in the upper half-space, which
means that $\langle v_n \rangle \simeq v_\ast$ for such contours, and
therefore, $\xi$ is given by the old formula~(\ref{xi_short}).

As one can see from Eqs.~(\ref{P}-\ref{P_2}) and~(\ref{P_approx}), our
approximate treatment does not capture the term $P_2(v, E)$ in $P(v,
E)$, which seems to be important for $v > v_\ast$. Therefore, the
possibility of $\min_n \langle v_n \rangle$ being larger than $v_\ast$
can not be totally ignored. Although we can not rigorously prove that
$\min_n \langle v_n \rangle = v_\ast$, we managed to find the {\em upper
bound\/} for $\min_n \langle v_n \rangle$,
\begin{equation}
       \min\limits_n \langle v_n \rangle < v_\ast + O(1)\, d,
\label{bound_on_v}
\end{equation}
based on the following percolation-type arguments.

Suppose that there exists a level line number $m$, $U_0[\rho_x^{(m)},
\rho_y] = E$, which is totally contained in a slab $\{(u, v, \rho_y), 0 \leq v
\leq v_0\}$. In this case $\min_n \langle v_n \rangle \leq \langle v_m
\rangle \leq v_0$. This prompts considering the following percolation
problem. Let us call ``wet'' all the points $(u, v, \rho_y)$ of the
slab, which satisfy the conditions
\[
{\rm Re}\, U_0(u - i v, \rho_y) \leq E,\quad
{\rm Im}\, U_0(u - i v, \rho_y) \leq 0.
\]
As the thickness $v_0$ of the slab increases from zero, it should
eventually reach a critical value $v_c$, at which the percolation
through ``wet'' regions first appears. The percolation threshold $v_c$
is at the same time the upper bound for $\min\limits_n \langle v_n
\rangle$. 

Obviously, no percolation exists for $v_0 < v_\ast$ when the wet regions
occupy exponentially small fraction of the volume. On the other hand, if
$v_0 \gg v_\ast$, then $\langle ({\rm Re}\, U_0)^2\rangle \simeq
\langle({\rm Im}\, U_0)^2 \rangle \simeq Q / 2 \gg E^2$, and
approximately a quarter of the entire volume is wet. This greatly
exceeds the volume fraction $0.17$ required for the percolation in
continual three-dimensional problems;~\cite{Isichenko} thus, such $v_0$
are high above the percolation threshold. Furthermore, the relation
between the percolation thresholds of the continuum and of a
film~\cite{Shklovskii_75} suggests that $v_c$ must be equal to $v_\ast +
C d$, where $C \sim 1$, which leads to Eq.~(\ref{bound_on_v}).

Finally, let us comment on the relation of our approach to that of Raikh
and Shahbazyan~\cite{Raikh_xi} already mentioned above. In
Ref.\ \onlinecite{Raikh_xi} the idea of the complex trajectories
$\rho_x^{(n)}(\rho_y) = u_n(\rho_y) - i v_n(\rho_y)$, which is the basis
of our calculation of the Green's function, was introduced. However, an
approximation early in their analysis [dropping of the
cross-product in Eq.~(2.8) of Ref.\ \onlinecite{Raikh_xi}] led the
authors of Ref.\ \onlinecite{Raikh_xi} to the result, which in our
notations can be written as follows:
\[
G_\rho(0, y; E) =  \frac{i}{l^2} \sum_n\,
[{\partial U_0}/{\partial\rho_x^{(n)}}]^{-1}
\exp\!\left[\frac{i}{l^2}\,y\,
\rho_x^{(n)}(y / 2)\right]
\]
(the derivative is taken at $\rho_y = y / 2$). As one can see, in their
method $G_\rho(0, y; E)$ is determined not by the entire tunneling
trajectory, as it should [see Eq.~(\ref{G_rho_WKB})], but only by its
midpoint $\rho_y = y / 2$. As a result, the point-to-point Green's
function $G(0, y; E)$ obtained by their method starts to be in error
strictly speaking already for $y \gtrsim d$. The only case where the
method of Raikh and Shahbazyan~\cite{Raikh_xi} works for arbitrary $y$
is the case of a one-dimensional potential, e.g., $U_0(x, y) = U_0(x)$.
A potential of this type has the same magnitude at the midpoint $y / 2$
and at all other points $y$. (From the formal point of view, only
for this type of potentials the aforementioned cross-product vanishes
for any $y$ and therefore can be dropped).

The reduction of the properties of the entire trajectory to the
properties of a single point destroys the self-averaging of the
localization length, which is a well established property of other
disordered systems.~\cite{Lifshitz} Because of that, it becomes
necessary to treat $\xi$ as a function of $y$. As we mentioned above,
for a {\it given disorder realization\/} the value of $G(0, y; E)$ given by
the formulas of Ref.\ \onlinecite{Raikh_xi} is in error already for $y
\gtrsim d$. However, this is not so for the quantity $\langle \ln | G(0,
y; E)| \rangle$. The {\it logarithmic averaging\/} manages to mask the error
so that $\xi(y)$ remains close to the correct asymptotic value given by
Eq.~(\ref{xi_short}) for sufficiently short distances, $y \ll d\,
\exp(E^2 / W_0^2)$. Indeed, if one uses the last equation above, then
one obtains
\[
|G|^2 \sim \sum_n \exp \left\{ -\frac{2}{l^2} \left[u_n^2(y / 2)
+ y\, v_n(y / 2)\right] \right\},
\]
instead of Eq.~(\ref{G_WKB II}) and
\begin{equation}
    \frac{1}{\xi(y)} \simeq  \frac{1}{l^2} \min \limits_n
\left[\frac{u_n^2(y / 2)}{y} + v_n(y / 2)\right]
\label{one_xi_RS}
\end{equation}
instead of Eq.~(\ref{one_xi}).  If $y \ll d\, \exp(E^2 / W_0^2)$, the
minimum is typically supplied by one of the trajectories whose ``height''
$v_n$ is not too different from the standard value of $v_\ast$, $|v_n -
v_\ast| \lesssim d$.  This range of $v_n$ allows for $u_n$ in the range
$|u_n| \lesssim \sqrt{y d}$.  Therefore, the optimal trajectory is
typically one of about $M = \sqrt{y / d}$ trajectories closest to the $x =
0$ axis. Note that $M$ increases with $y$.  Eventually, at $y \sim d\,
\exp(E^2 / W_0^2)$ it becomes exponentially large so that there is an
appreciable probability that one out of such $M$ trajectories has $v_n(y /
2) < v_\ast / 2$ and $|u_n(y / 2)|^2 < y d / 2$.  The decay rate $1 /
\xi(y)$ in this case is determined by such an untypical trajectory, and so
$\xi(y)$ is significantly larger than our result, Eq.~(\ref{xi_short}).
Furthermore, at $y \gg d\, \exp(E^2 / W_0^2)$, there is a finite probability
of finding $v_n(y / 2)$ exactly equal to zero [due to the third term in
Eq.~(\ref{P})]. Therefore, in the asymptotic limit $y \to \infty$ the
method of Raikh and Shahbazyan~\cite{Raikh_xi} yields a very surprising result
$1 / \xi(y) \to 0$.  As we explained above, this is a consequence of an
effective substitution of the original two-dimensional random potential by a
potential depending on a single coordinate.

\section{Overlapping Landau levels}
\label{percolation}

This section is devoted to the derivation of Eq.~(\ref{xi_milnikov}). An
important difference from all the preceding calculations is that the
potential energy $E = E_F - \hbar\omega_c(N + \frac12)$ is smaller than
the amplitude $W_0$ of the averaged potential $U_0(\bbox{\rho})$. In
this case the level lines $U_0[\rho_x^{(n)}, \rho_y] = E$ introduced in
Sec.~\ref{long}, stay predominantly in the real plane $v = 0$. [Their
density is given by $P_3(v, E)$, see Eq.~(\ref{P})]. However, for $E
\neq 0$ the percolation of the level lines in the $y$-direction
discussed in the end of the previous section, still requires brief
excursions into the upper complex half-space. Such excursion link
together the large closed loop contours $U_0 = E$ of typical diameter
$\xi_{\rm perc}(E)$. The localization length $\xi$ is still given by
Eq.~(\ref{xi_v}), which leads to the estimate
\begin{equation}
\xi \sim \frac{l^2}{\langle v \rangle_{\rm ex}}\,
         \frac{\xi_{\rm perc}(E)}{\Delta\rho}, 
\label{xi_percolation}
\end{equation}
where the subscript ``ex'' indicates that the averaging is performed
only over the ``excursions'', i.e., over the parts with $v > 0$, with
$\Delta \rho$ being the typical length of such parts. Similar to
the case $N = 0$ discussed previously by Mil'nikov and
Sokolov,~\cite{Milnikov} $\langle v \rangle_{\rm ex} \sim \Delta\rho
\sim (|E| / U_0'')^{1/2}$ where $U_0'' \sim W_0 / d^2$ is the typical
value of the second derivatives of the averaged potential $U_0$;
therefore,
\begin{equation}
          \xi \sim \xi_{\rm perc}(E)\, \frac{W_0 l^2}{E d^2}.
\label{xi_percolation II}
\end{equation}
The calculation of the quantity $\xi_{\rm perc}(E)$ is a subject of
statistical topography,~\cite{Isichenko} and the following results
are established. Denote by $C_\lambda$ the integral,
\[
C_\lambda \equiv \int\limits_{\frac12 < |q| \lambda < 1}\!\!\!\!\!
\frac{d^2 q}{(2\pi)^2}\,
\tilde{C}_0(q).
\]
For slowly decaying correlators $\tilde{C}_0(q)$, such that $C_\lambda \sim
\lambda^{-2 H}$ with $H \leq 3 / 4$, $\xi_{\rm perc}$ is given
by (besides the review, Ref.~\onlinecite{Isichenko},
see the original works, Ref.~\onlinecite{Weinrib}),
\begin{equation}
  \xi_{\rm perc}(E) \sim d\, |W_0 / E|^{1 / H}.
\label{Weinrib exponent}
\end{equation}
Otherwise, i.e., if $C_\lambda$ decays faster than $\lambda^{-3 / 2}$,
then~\cite{Isichenko}
\begin{equation}
  \xi_{\rm perc}(E) \sim d\, |W_0 / E|^{4/3}.
\label{standard exponent}
\end{equation}
%

%
%
\begin{figure}
\centerline{
\psfig{file=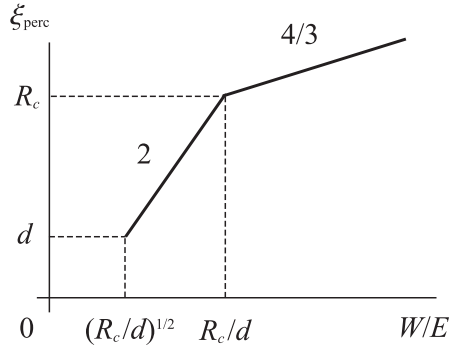,width=2.3in,bbllx=208pt,bblly=401pt,bburx=502pt,bbury=633pt}
}
\vspace{0.1in}
\setlength{\columnwidth}{3.2in}
\centerline{\caption{
The percolation length $\xi_{\rm perc}$ as a function of the energy $E$
(schematically). The labels ``2'' and ``4/3'' indicate the power-law
exponents in the corresponding intervals, see Eq.~(\protect\ref{xi_perc}).
\label{Plot_xi_perc}
}}
\end{figure}

Let us now determine the conditions under which formulas~(\ref{Weinrib
exponent}) and~(\ref{standard exponent}) become applicable. The issue is
complicated by the fact that $\tilde{C}_0(q)$ is the product of two
terms, $\tilde{C}_0(q)$ and $[F_N(q l^2)]^2$, see Eq.~(\ref{C_0}). The
former decays exponentially starting from $q \sim 1 / d$. The latter
remains close to one at $q \lesssim 1 / R_c$, then behaves according to a
power-law, $[F_N(q l^2)]^2 \sim 1 / q$ for $1 / R_c < q < 2 k_F$, and
finally, decays exponentially at $q > 2 k_F$. Such a complicated behavior
results into three different regimes~(\ref{xi_milnikov}a-c).

The simplest is the case $R_c \ll d$, where $[F_N(q l^2)]^2 \simeq 1$
for all relevant $q$. In this case Eq.~(\ref{standard exponent}) applies
and also $W_0 \simeq W$. The localization length is given by
Eq.~(\ref{xi_milnikov}a), which coincides with the result of Mil'nikov
and Sokolov.~\cite{Milnikov} 

If $R_c \gg d$, then the situation is more complicated. In this case
$C_\lambda$ is proportional to $1 / \lambda$, i.e., $H = \frac12$ for $d
\ll \lambda \ll R_c$, yet decays faster than $1 / \lambda^{3/2}$ for
$\lambda > R_c$. As a result, both Eq.~(\ref{Weinrib exponent}) and
Eq.~(\ref{standard exponent}) for $\xi_{\rm perc}$ may apply, depending
on $E$,
\begin{equation}
\xi_{\rm perc}(E) \sim \left\{
\begin{array}{lc}
\displaystyle \frac{d^2}{R_c} \left|\frac{W}{E}\right|^2,
&\displaystyle \frac{d}{R_c} < \frac{|E|}{W} < \sqrt{\frac{d}{R_c}},\\[12pt]
\displaystyle R_c\left|\frac{d}{R_c}\frac{W}{E}\right|^{\frac43},
&\displaystyle  |E| < W\frac{d}{R_c},
\end{array}
\right.
\label{xi_perc}
\end{equation}
which is illustrated by Fig.~\ref{Plot_xi_perc}. Combining
Eq.~(\ref{xi_percolation II}) where $W_0 \sim W \sqrt{d / R_c}$ with
Eq.~(\ref{xi_perc}) and using $|E| = \hbar\omega_c / 2$, one obtains
Eqs.~(\ref{xi_milnikov}b) and~(\ref{xi_milnikov}c).

\section{Discussion of a more realistic model and
its comparison with the experiment}
\label{conclusions}

To make the connection with the experimental practice we will consider
the model where the random potential is created by randomly positioned
ionized donors with two-dimensional density $n_i$ set back from the
electron gas by an undoped layer of width $d$. We will assume that $1 /
d^2 \ll n_i \ll n^2 d^2$. In zero magnetic field the random potential can
be considered a weak Gaussian random potential with the correlator
\begin{equation}
         \tilde{C}(q) = \pi^2 n_i (e^2 a_B)^2\, e^{-2 q d},
\label{C_Coulomb}
\end{equation}
where $a_B = \hbar^2 \kappa / m e^2$ is the effective Bohr radius (see
Appendix B of Ref.~\onlinecite{Fogler_chaos}). This formula corresponds
to $\beta = 1$ and $d$ replaced by $2 d$ in Eq.~(\ref{C_q}). Deriving
Eq.~(\ref{C_Coulomb}), we took into account the screening of the donors'
potential by the electron gas described by the dielectric function~\cite{Ando}
\begin{equation}
\varepsilon(q) = \kappa \left(1 + \frac{2}{a_{\rm B} q}\right),
\quad q \leq 2 k_F.
\label{epsilon_TF}
\end{equation}
This model remains to be accurate in sufficiently weak fields, where the
Landau levels overlap and the density of states is almost uniform, like
in zero field. In stronger fields (the boot-shaped region AHIGEB in
Fig.~\ref{diagram}), the density of states develops sharp peaks at the
Landau level centers separated by deep minima. This strongly influences
the property of the electron gas to screen the external impurity
potential. Different aspects of such a screening in {\it weak\/}
magnetic fields have been addressed in
Refs.~\onlinecite{Kukushkin,Aleiner,Fogler_nonlinear}.
The screening can be be both linear and nonlinear, depending on the wave
vector $q$.

The concept of nonlinear screening has been developed by Shklovskii and
Efros initially for the three-dimensional case.~\cite{ES} Gergel and
Suris~\cite{Gergel} have extended it to the two-dimensional case. The
influence of a strong magnetic field on the nonlinear screening has been
studied in Ref.~\onlinecite{Shklovskii_86} and especially in
Ref.~\onlinecite{Efros}.

Nonlinear screening is realized for sufficiently small $q$, $q < q_{\rm
nonl}$, and is enhanced compared to the zero field case
[Eq.~(\ref{epsilon_TF})]. The threshold wave vector $q_{\rm nonl}$ is a
complicated function of the magnetic field.~\cite{Fogler_nonlinear} If
$q_{\rm nonl}$ is smaller than $1 / R_c$, there exists an intermediate
range of $q$, $q_{\rm nonl} < q < 1 / R_c$, where the screening remains
in the linear regime but is suppressed compared to
Eq.~(\ref{epsilon_TF}). The corresponding dielectric function is given
by~\cite{Kukushkin,Aleiner}
\begin{equation}
\varepsilon(q) \simeq \kappa \left(1 + \frac{R_c^2 q}{a_{\rm B}}\right).
\label{epsilon}
\end{equation}
At even larger $q$, $q > \max\{q_{\rm nonl}, 1 / R_c\}$, there is no
essential change in the screening properties brought about by the magnetic field.

Clearly, in such magnetic fields the model of a Gaussian random
potential with the correlator~(\ref{C_Coulomb}) becomes an
oversimplification. Fortunately, the localization length $\xi$ should
not be strongly affected by this. Indeed, $\xi$ is sensitive only to the
combined amplitude of the narrow band of harmonics with wave vectors $q
\simeq q_\ast$. It can be shown that for $\beta$ equal to one, $q_\ast \simeq 2
k_F$ (just like for the ``white-noise'' potential). This wave vector
belongs to the last group of $q$ for which there is no change in the
screening properties. In fact, at such $q$ the screening is equally
ineffective both in zero and in arbitrary strong magnetic field,
$\varepsilon(2 k_F) \simeq \kappa$.

The parameter $k_F d$ is almost always larger than one in the
experiment; therefore, in strong fields $\xi$ is expected to be given by
Eq.~(\ref{xi_short}). For $\beta = 1$ this formula becomes remarkably
simple,
\[
  \xi \simeq \frac{l^2}{d} = \frac{R_c}{k_F d}.
\]
The simplest way to obtain this expression is to start with
Eq.~(\ref{xi_short}) for $\beta > 1$, take the limit $\beta \to 1$, and
then make a replacement $d \to 2 d$ (see above).
For $k_F d$ close to unity, however, one should use a refined formula
\begin{equation}
  \xi \simeq \frac{R_c}{k_F d + \frac12 \ln(\hbar\omega_c / W_0)},
\label{xi_coulomb}
\end{equation}
which follows from the master equation~(\ref{xi_master}).
Formula~(\ref{xi_coulomb}) should hold in the boot-shaped region AHIGEB
of Fig.~\ref{diagram}. The corresponding range of $N$ can be expressed
in terms of the dimensionless parameters $k_F d$ and $n_i / n$. We will
concentrate on the case $n_i \sim n$, which we call ``standard.'' In
this case Eq.~(\ref{xi_coulomb}) is valid for $N \lesssim 2 k_F d$. At
larger $N$, the localization length is given by Eq.~(\ref{xi_milnikov})
so that the dependence of $\xi$ on $\nu$ is super-linear. Note that the
``standard'' case corresponds to the straight-line ``trajectory''
passing through the points E and F and shown by the arrows in
Fig.~\ref{diagram}. Between these points, Eq.~(\ref{xi_milnikov})
reduces to
\begin{equation}
\frac{\xi}{R_c} \sim \frac{N^{3 / 2}}{(k_F d)^{5 / 2}}, \quad
2 k_F d \lesssim N \lesssim C (k_F d)^{5/3},
\label{xi_EF}
\end{equation}
where $C$ is some undetermined numerical factor. At even larger $N$,
$\xi$ becomes exponentially large, which precludes its accurate
measurement at experimentally accessible temperatures. For this reason we
do not give the explicit formula for the localization length at such $N$.

Unfortunately, the published experimental data on the low-temperature
magnetoresistance away from the QHE peaks is limited to the measurements
done by Ebert {\it et al.\/}~\cite{Ebert} more than a decade ago.  In most
of their samples $d$ was equal to 6 {\rm nm}, which corresponds to $k_F d
= 0.9$. On the one hand, this places the samples close to the line HI in
Fig.~\ref{diagram}.  On the other hand, the ``standard'' case corresponds to
the line EF.  In fact, there is no contradiction here because in the
``standard'' case $E_F / W \sim k_F d$.  Therefore, the as $k_F d$ tends to
unity, the boot-shaped region AHIGEB shrinks and the lines EF and HI become
quite close.

With the help of Eq.~(\ref{T_0}), we converted $T_0$
reported by Ebert {\it et al.\/}~\cite{Ebert} into $\xi$. Only even
$\nu$ were selected for this analysis. The constant factor in
Eq.~(\ref{T_0}) was chosen to be $6.0$ following Nguyen.~\cite{Nguyen}
For filling factors $\nu = 2$ and $\nu = 4$, i.e., $N = 1$ and $N = 2$,
there is a good agreement with Eq.~(\ref{xi_coulomb}) if the
logarithm in the denominator is neglected. At larger $\nu$,
$6 \leq \nu \leq 12$, empirical $\xi$ behaves roughly like
\begin{equation}
                \xi \approx 2.2 N^{3/2} R_c.
\label{Ebert_fit}
\end{equation}
Formally, this is in agreement with Eq.~(\ref{xi_EF}). However, since the
parameter $k_F d$ is close to one, the random potential can not be treated
as long-range, and so the percolation picture used in derivation of
Eq.~(\ref{xi_EF}) is not quite justified.  Perhaps, it is more resonable to
treat Eq.~(\ref{Ebert_fit}) as an approximation to the exponential
dependence, Eq.~(\ref{ansatz}), within a limited range of $\nu$.

Another comment is in order here. Formula~(\ref{T_0}) is based on
the assumption that the interaction energy of two quasiparticles
separated by a typical hopping distance $r$ is given by the Coulomb law.
As was shown in Ref.~\onlinecite{Aleiner}, with a dielectric function
given by Eq.~(\ref{epsilon}), the Coulomb law is realized only at sufficiently
large distances, $r \gg R_c^2 / a_B$. Large hopping distances $r$
correspond to low temperatures, $T = (\xi / 4 r)^2 T_0$. Therefore,
formula~(\ref{T_0}) is expected to hold only for
\begin{equation}
             T < \left(\frac{a_B \xi}{4 R_c^2}\right)^2 T_0.
\label{no_screening}
\end{equation}
Experimentally, such low temperatures may be hard to reach, especially in
weak fields where $R_c$ is large. Therefore, we think that a different type
of measurement may be more promising. One has to measure both the
temperature and the current dependence of the magnetoresistance.  Comparing
the results of these two types of measurements, one can determine the {\it
effective temperature\/} $T_{\rm eff}$ for each value of the current density
$j$.  As discussed in Ref.~\onlinecite{Polyakov}, the relation between
$T_{\rm eff}$ and $j$ is as follows,
\begin{equation}
           k_B\, T_{\rm eff}(j) \simeq 0.5\, e j \rho_{xy}\, \xi,
\label{T_eff}
\end{equation}
where $\rho_{xy}$ is the Hall resistivity. Equation~(\ref{T_eff}) does
not involve the dielectric function and therefore, is expected to hold
even when Eq.~(\ref{T_one_half}) and the corresponding current
dependence~\cite{Polyakov} $\rho_{xx} \propto e^{-\sqrt{T_0 / T_{\rm
eff}(j)}}$ do not match the magnetoresistance data exactly.


\acknowledgements

We are grateful to V.~I.~Perel for a valuable contribution to this work,
to M.~E.~Raikh and T.~V.~Shahbazyan for useful comments, to
V.~J.~Goldman and L.~P.~Rokhinson for sharing with us their unpublished
data, and to A.~A.~Koulakov for a critical reading of the manuscript.

A.~Yu.~D. is supported by the Russian Foundation for the Basic
Research. M.~M.~F. is a recipient of University of Minnesota's Doctoral
Dissertation Fellowship. M.~M.~F. and B.~I.~S. acknowledge
support from NSF under Grant DMR-9616880.
\appendix
\section{Level line density $P(\lowercase{v}, E)$}
\label{P_appendix}

The equation $U_0 = E$ is equivalent to the system of two equations $R = E$
and $I = 0$, where $R = {\rm Re}\,U_0(u - i v, \rho_y)$ and $I = -{\rm
Im}\,U_0(u - i v, \rho_y)$.  To calculate $P(v, E)$ we need to know the
Jacobian $J = |\partial (R, I) / \partial(u, v)|$.  With the help of the
Cauchy relations,
\[
   \partial R / \partial v = \partial I / \partial u,\:\:\:
   \partial R / \partial u = -\partial I / \partial v,
\]
the Jacobian can be written as
\[
                       J = R_v^2 + I_v^2,
\]
where subscripts denote the partial derivatives. Therefore, for all $v \neq 0$,
\begin{eqnarray}
\displaystyle P(v, E) &=& \int\!
\frac{d R\,d R_v\,d I\,d I_v}{(2 \pi)^2 (\det {\sf K})^{1/2}}\,
\delta(R - E)\, \delta(I)\, (R_v^2 + I_v^2)\nonumber\\
\displaystyle \mbox{} &\times& \exp
\left(-\frac12 {\bf v}^T {\sf K}^{-1} {\bf v} \right),
\label{P_integral}
\end{eqnarray}
where ${\bf v}$ is the four-component vector $(R, R_v, I, I_v)^T$ and ${\sf
K}$ is its autocorrelation matrix, ${\sf K}_{m n} = \langle v_m v_n
\rangle$.  Matrix ${\sf K}$ turns out to be block-diagonal, the blocks being
$2 \times 2$ matrices.  Matrix elements of ${\sf K}$ can be calculated using
Eq.~(\ref{C_r}).  For example, the upper diagonal block has elements ${\sf
K}_{1 1} = Q / 2$, ${\sf K}_{1 2} = {\sf K}_{2 1} = S / 2$, and ${\sf K}_{2
2} = R / 2$ [see Eq.~(\ref{QRST}) for definitions]. The subsequent
integration over $R_v$ and $I_v$ in formula~(\ref{P_integral}) becomes
a trivial task and yields the result represented
by Eqs.~(\ref{P_1}) and~(\ref{P_2}).

The case $v = 0$ requires a special consideration because the condition $I =
0$ is satisfied identically.  It is easy to see that a level line would
typically branch upon the intersection with the $v = 0$ plane.  The
additional branch or branches are totally contained in this plane, which
gives the delta-function-like contribution to function $P(v, E)$.  The
coefficient in front of the delta-function is given by
\begin{eqnarray*}
\displaystyle P_3(E) &=& \int\!
\frac{d R\,d I_v}{2 \pi ({\sf K}_{11} {\sf K}_{44})^{1/2}}\,
\delta(R - E)\, |I_v|\\
\displaystyle \mbox{} &\times& \exp
\left[-(R^2 / 2 {\sf K}_{11}) - (I_v^2 / 2 {\sf K}_{44}) \right]\\
&=& \frac{1}{4\pi} \sqrt{-\frac{{\nabla}^2 C_0(0)}{C_0(0)}}\,
e^{-E^2 / 2 W_0^2}.
\end{eqnarray*}
As expected, $P_3(E)$ is exponentially small for $|E| \gg W_0$.


\end{multicols}

\begin{references}
\bibitem{Comment on Ward} Note that the quantity $\langle G(0, y, E + \omega)
G^*(0, y, E - \omega)\rangle$ diverges like $1/\omega$ as $\omega\to 0$.
Therefore, the average square modulus of the {\em exact\/} Green's function is
infinite. The quantity we calculate, $\langle |G(0, y, E)|^2\rangle_{\rm nr}$
is perfectly finite and can be used to get meaningful results for $\xi$.
\end{references}
\end{document}